\documentclass[lettersize, journal]{IEEEtran}
\PassOptionsToPackage{table}{xcolor}
\ifCLASSOPTIONcompsoc
  \usepackage[nocompress]{cite}
\else
  \usepackage{cite}
\fi

\usepackage{subcaption}
\usepackage{graphicx}
\usepackage[hidelinks]{hyperref}
\usepackage{booktabs}
\usepackage{multirow}
\usepackage{xcolor}
\usepackage{tcolorbox}

\begin{document}

\title{Adapting Installation Instructions in Rapidly Evolving Software Ecosystems}

\author{Haoyu~Gao,
        Christoph~Treude,
       Mansooreh~Zahedi}

\markboth{Journal of \LaTeX\ Class Files,~Vol.~14, No.~8, August~2015}%
{Shell \MakeLowercase{\textit{et al.}}: Bare Demo of IEEEtran.cls for IEEE Journals}

\maketitle

\begin{abstract}
README files play an important role in providing installation-related instructions to software users and are widely used in open source software systems on platforms such as GitHub. Software projects evolve rapidly alongside their dependencies in dynamic software ecosystems, requiring frequent updates to installation instructions. These instructions are crucial for users to start with a software project. Despite their significance, there is a lack of systematic understanding regarding the documentation efforts invested in README files and the triggers behind them.
To fill the research gap, we conducted a qualitative study, investigating 400 GitHub repositories with 1,163 README commits that focused on updates in installation-related sections. Our research revealed six major categories of changes in the README commits, namely pre-installation instructions, installation instructions, post-installation instructions, help information updates, document presentation, and external resource management. We further provide detailed insights into modification behaviours and offer examples of these updates. We also studied the triggers for the documentation updates, which led to three categories including errors in the previous documentation, changes in the codebase, and need for documentation improvement. Based on our findings, we proposed a README template tailored to cover the installation-related sections for documentation maintainers to reference when updating documents. We further validated this template by conducting an online survey and a pull request study, identifying that documentation readers find the augmented documents based on our template to be generally of better quality, and documentation maintainers find it useful. We further provide recommendations to practitioners for maintaining their README files, as well as motivations for future research directions. These recommendations encompass completeness, correctness and up-to-dateness, and information presentation considerations. The proposed research directions include the development of automated tools, in particular for documentation updates, and conducting empirical studies to enhance comprehension of the needs of documentation users. 

\end{abstract}


\section{Introduction}
Good software documentation serves as a vital source of information for different stakeholders involved in software systems. In particular, README files are among the first software documents that users, such as other developers and end-users, and newcomers encounter when accessing an Open Source Software (OSS) project, leaving them with a crucial first impression and providing them with an introductory overview of the project~\cite{prana2019categorizing, koskela2018open}. At the time of this study, GitHub, the most popular open-source code platform, hosts more than 372 million repositories,\footnote{\url{http://web.archive.org/web/20230424004513/https://en.wikipedia.org/wiki/GitHub}} underscoring the high demand for maintaining high-quality README files.

One of the most common types of content in GitHub README files is related to basic instructions (e.g., project installation guide)~\cite{prana2019categorizing}.  A well-written ``How'' section could alleviate the effort for developers and users, especially newcomers, to get hands on with the projects. In fact, overcoming the challenge of setting up an environment is one of the hurdles for newcomers to contribute to OSS projects~\cite{steinmacher2014preliminary}.  

Software projects evolve rapidly alongside their dependencies in dynamic software ecosystems~\cite{boucharas2009formalizing}, requiring frequent updates to installation instructions. Therefore, README files need to be carefully maintained; otherwise, various software documentation issues may arise. For example, software tutorials that focus on installing software tools, configuring virtual computing environments, and other related continuous deployment processes have been found to be not executable in more than half of the cases investigated by related work~\cite{mirhosseini2020docable}. Furthermore, previous empirical studies identified inappropriate installation instructions, outdated content, and content clarity as frequently appearing documentation issues~\cite{aghajani2019software,aghajani2020software}. These factors collectively constitute the impediments for both developers and project users to use software repositories, while empirical evidence on the current documentation practices and their triggers could pave the way for mitigating these issues from source.

Although previous studies have pinpointed software documentation issues, particularly in installation-related topics~\cite{aghajani2019software, aghajani2020software, vskoric2016selecting}, their analyses were broad in scope, with the aim of synthesising different aspects of documentation issues from surveys or software documents. In contrast, documentation updates provide a new perspective consisting of the modifications and two states: before and after the modification, while triggers offer the motivations that drive documentation updates in the rapidly evolving software ecosystem. In fact, increased documentation activities and efforts have demonstrated their efficacy in improving the popularity of and encouraging participation in GitHub projects~\cite{aggarwal2014co}. Considering the documentation issues and the benefits of investing efforts to maintain high-quality documentation, knowledge of the practices of the README file efforts and the triggers behind them becomes essential. This knowledge reveals documentation activities in the dynamic software ecosystems.
It not only provides practitioners with recommendations but also provides researchers with insights for the development of automated documentation tools. However, empirical evidence to support practices related to software documentation efforts remains scarce.

In this work, we conducted an empirical study to qualitatively analyse README file modification activities in installation-related topics to fill the research gap. We mined GitHub repositories from GHArchive that were active between 2015-01-01 and 2023-02-28,\footnote{\url{https://www.gharchive.org/}} applying stringent filters on different features for the repositories, resulting in 12,908 projects. We examined a statistically significant sample of 1,168 README file commits within 400 randomly selected repositories and adopted qualitative analysis to synthesise the types of README update tasks in the modified sections. 

Based on our analysis, we synthesised a comprehensive taxonomy of 189 update behaviours, which are related to the categories of: \textit{(1) pre-installation instructions, (2) installation instructions, (3) post-installation instructions, (4) help information, (5) document presentation}, and \textit{(6) external resource management}. We then devised a process to identify the triggers for documentation updates. Our analysis leads to three categories of documentation update triggers: \textit{(1) errors in the previous documentation, (2) changes in the codebase, and (3) need for documentation improvement.}  After that, we proposed a README template tailored to cover the installation-related sections for documentation maintainers to reference when updating documents. The template validation process indicates that readers find README files that apply our template to be of generally increased quality, and the documentation maintainers also find the process of applying the template useful. This serves as a fundamental step towards automated documentation updates. We further discussed their implications for documentation maintainers and software engineering researchers based on our discoveries from perspectives of completeness, correctness and up-to-dateness, and information presentation considerations. 

Our key contributions are as follows: (1) A large-scale repository dataset, which can be used for future research. (2) Formalisation of a detailed list of software development tasks related to README file content. (3) A comprehensive taxonomy for the types of installation-related documentation modifications for OSS projects. (4) A README template validated by documentation readers to be of better quality compared to original README files. (5) A categorisation of triggers for documentation updates (6) Actionable recommendations for OSS documentation maintainers and researchers, in particular the automated documentation updates direction.

The rest of the paper is structured as follows: Section 2 covers the related work; Section 3 introduces our study design and methodology adopted to conduct our study; Section 4 presents the results of our qualitative analysis for the installation instruction taxonomy; Section 5 presents the results for triggers for documentation updates; Section 6 provides the README template from our study and its validation process with an online survey; Section 7 discusses the implications of this study to practitioners and researchers; Section 8 covers the threats to the validity of our study; Section 9 provides the conclusion. Our replication package is at \url{https://github.com/Haoyu-Gao/README_Empirical_Study}, with the dataset in the root folder.


\section{Related Work}

\subsection{Empirical Studies on Software Documentation}
Various empirical studies have been conducted to cover software documentation from different perspectives. 

One source of information for researchers is directly from software engineering practitioners. Several studies have investigated the impact of software documentation.  Kajiko-Mattson~\cite{kajko2005survey} studied the practices of software documentation in Swedish organisations, identifying rudimentary documentation requirements relevant within corrective maintenance. Garousi et al.~\cite{garousi2015usage} conducted an industrial case study on the usage and usefulness of different documentation artefacts in the software development life cycle. Their findings indicated variations in documentation usage based on information needs, and the recommendations derived from their study have been integrated into the company's practices where the experiment took place. Software documentation issues are also investigated. Uddin and Robillard~\cite{uddin2015api} conducted a large-scale survey on API documentation failure factors among IBM practitioners, and identified the most pressing issues related to content, as opposed to presentation. Aghajani et al.~\cite{aghajani2020software} presented the views of practitioners on various issues of software documentation, along with their proposed solutions to address these issues. Moreover, the quality of software documents is another perspective that has been studied. Treude et al.~\cite{treude2020beyond} recruited technical editors and synthesised a framework to assess the quality of the software documentation from the respondents. Their work provides visions for a potential unified quality framework for assessing software documentation.

However, these studies are limited in their context (e.g., practitioners' organisation and knowledge or software artefacts), making it difficult to generalise the results to broader software documentation artefacts. Another source of information for researchers lies in these software artefacts themselves. Different software documentation artefacts have been evaluated, revealing a series of issues. Wen et al.~\cite{wen2019large} 
conducted a large-scale empirical study of code-comment inconsistencies, identifying the taxonomy of inconsistencies fixed by developers. Mirhosseini and Parnin~\cite{mirhosseini2020docable} evaluated the executability of software tutorials and discussed the categories of errors that can lead to the failure of code block executions. Aghajani et al.~\cite{aghajani2019software} conducted a qualitative study on software documentation issues from Stack Overflow, GitHub issues, GitHub pull requests, and mailing lists, identifying both content and process issues. Gao et al.~\cite{gao2024documenting} studied machine learning documentation to highlight the insufficient effort in documenting ethical considerations.

Among the software documentation artefacts, README files play an important role in introducing OSS repositories. They are relatively easy to access, with a rich content editing history. Despite their significance, README files have received relatively less attention in research, and most existing work focuses on more abstract perspectives, such as structure and content topics.  Prana et al.~\cite{prana2019categorizing} conducted a qualitative study involving manual annotation of 4,226 README file sections to categorise README content into different purposes. Liu et al.~\cite{liu2022readme} studied the patterns adopted for README files by GitHub Java repositories and investigated the association between repository popularity and README file structures. The study identified 32 clusters representing common README file structures, and the findings suggested that repositories adhering to GitHub guidelines for README files tend to get more stars. Meanwhile, Wang et al.~\cite{wang2023study} investigated the factors that influence the popularity of GitHub repositories from the README file angle, and identified a significant difference between popular and non-popular repositories in README file-related features.

The richness of README file edit history provides us with the possibility to investigate software documents from a process perspective. In this work, we provide a more detailed study of README files centred around installation-related instructions, considering both their content and their process, to offer concrete suggestions for documentation maintainers and researchers.

\subsection{Automated Tools for Software Documentation}
Inspired by various empirical pieces of evidence regarding the roles and issues of software documentation, there has been much research on automated tool support for software documentation.

Previous research focuses on supporting various software documentation artefacts with non-documentation artefacts, such as source code. To begin with, code summarisation is an actively evolving field, with techniques developed for generating code comments~\cite{fowkes2017autofolding, hu2018deep, hu2020deep, wang2020reinforcement}, commit messages~\cite{xu2019commit, liu2020atom, dong2022fira}, code reviews~\cite{shi2019automatic, lin2024improving} and release notes~\cite{moreno2014automatic, moreno2016arena}. Technical debt, including code comment inconsistency detection~\cite{panthaplackel2021deep, liu2021just}, and self-admitted technical debt~\cite{wattanakriengkrai2019automatic, li2022self}, is also investigated to facilitate software maintenance.

Efforts have also been made to improve and augment software documentation content and navigation across this content. Treude and Robillard~\cite{treude2016augmenting} investigated a technique for augmenting the content of API documents by incorporating information from Stack Overflow. Treude et al.~\cite{treude2014extracting} extracted development tasks from software documents, bridging the gap between software documentation and the information needs of software developers. Gao et al.~\cite{gao2023evaluating} mined README files from GitHub and performed transfer learning to simplify complicated documentation sentences. In fact, Robillard et al.~\cite{robillard2017demand} have proposed a paradigm shift for automatic documentation generation based on developers' queries, given their contexts. 

However, to facilitate more effective automated documentation-writing tools, it is essential to have a deeper understanding not only of the software documentation artefacts but also of the patterns of behaviours associated with their updates. In this work, we conducted a qualitative analysis of installation-related documentation update behaviours along with the triggers behind them, with the aim of gaining a deep understanding and providing motivation for developing automated software tools for documentation writing.

\subsection{OSS Barriers and Project Installation}
The term ``installation'' is widely used in different contexts. A simple installation procedure only requires fetching a remote binary file, while a complicated situation would confirm that minimum system requirements are met, identify and resolve
shared library dependencies, verify the integrity of each downloaded component, and other procedures~\cite{barrera2014securing}. \textit{Due to the absence of a precise definition of installation-related instructions, we define them to encompass all procedures occurring after the environment setup and the preceding project usage phase in this paper.}

Recent studies~\cite{steinmacher2014preliminary, steinmacher2015systematic} have highlighted both documentation problems and the configuration of the development environment as barriers for newcomers when they first contribute to Open Source Software (OSS) repositories. Salerno et al.~\cite{salerno2023barriers} discovered that student self-efficacy improves as a result of participating in OSS courses, but simultaneously they also perceive an increase in documentation issues. In particular, one of the identified issues revolves around the challenges associated with setting up the development environment.

Moreover, another study~\cite{prana2019categorizing} pointed out that the major content category of README files is ``How'', which includes instructions from setting up the environment to installation and running the project. Therefore, README files play a vital role in guiding newcomers and software users with installation-related topics. 

In contrast to previous research on the human perspective of OSS barriers for newcomers, this work synthesises empirical evidence on software documentation activities, specifically on parts related to the installation, based on evidence derived from a large-scale dataset that we mined. Such empirical evidence will be useful to mitigate documentation-related barriers for newcomers when engaging with OSS projects.

\section{Study Design and Methodology}
The goal of this study is to investigate how installation-related instructions in README files are updated. We formulate three research questions (RQs) to guide our study:

\textbf{RQ1: How are installation sections in README files modified?} This RQ will reveal different modification behaviours for updating the documentation. We studied the documentation modification behaviour rather than the most recent version of README files because the modification behaviours not only provide the documentation structure and contents but also offer more insights into the documentation maintenance efforts on OSS repositories. By understanding the documentation maintenance practice, we could provide implications to practitioners and researchers for better maintaining the README files. We performed a qualitative analysis of the installation-related sections within our mined data to answer this question. One of the target outcomes of this RQ is a template encompassing all essential components to facilitate better documentation activities for the maintainers.

\textbf{RQ2: What are the triggers for the README updates in the installation sections?} This RQ investigates the reasons that trigger documentation updates in the software system, providing insights into what drives documentation updates in a rapidly evolving software ecosystem. The triggers and update behaviours collectively provide strong empirical evidence for automatic documentation updates as a future direction.

\textbf{RQ3: Could a template for installation instructions be devised to improve documentation quality?} This research question aims to validate our proposed README template, which is the fundamental first step towards automated documentation updates. We conducted a survey asking participants to evaluate the perceived documentation quality before and after applying the README template, based on an established documentation assessment framework~\cite{treude2020beyond}. We further conducted a pull request study, similar to Malaquias et al.~\cite{malaquias2017discipline} to evaluate the practitioners' perspective on our template.

\subsection{Data Preparation}

In this paper, we only consider GitHub repositories that use Python or Java as their main programming language. Python and Java are two widely used programming languages that span a broad range of applications, from developing easy-to-use tools to large-scale servers. In fact, Python and Java consistently rank among the top three most used programming languages on GitHub,\footnote{\url{https://octoverse.github.com/2022/top-programming-language}} following JavaScript.  

\begin{figure}[t!]
    \centering
    \includegraphics[width=\columnwidth]{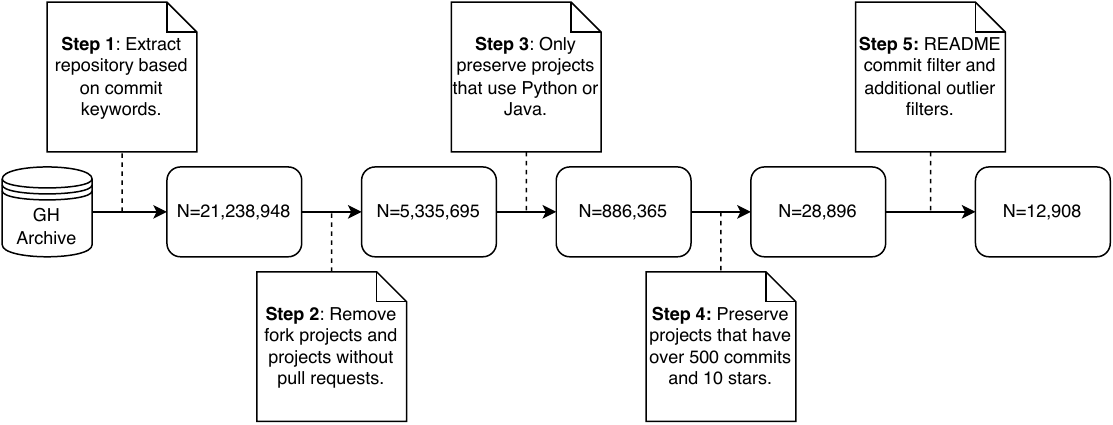}
    \caption{Data Preparation Process}
    \label{fig:data preparation}
\end{figure}

Although the GitHub API provides access to its data, the rate limit is not enough for collecting all repositories that meet our needs. Therefore, we used GH Archive,\footnote{\url{https://www.gharchive.org/}} which collects GitHub data based on the Event API. Figure~\ref{fig:data preparation} describes our overall repository collection procedure.

As our focus is on update behaviours, we started with commits rather than repository content. In addition, we want the repositories retrieved to contain a history of updating their installation-related instructions. Therefore, we looked at the ``PushEvent'' in the data records, and retrieved those with commit messages containing keywords that we pre-defined after tokenisation and lemmatisation. Keywords were derived from four base keywords ``instruct'', ``guide'', ``install'' and ``deploy''. These terms were chosen to represent the content genres and topics that we aimed to include in our analysis.  They were further combined with tags used in Zahedi et al.~\cite{zahedi2020mining}. We expand the set of keywords by finding both the noun and verb forms of words in WordNet~\cite{miller1995wordnet}. This procedure resulted in a total of 21,238,948 repositories. After that, we preserved only 5,335,695 repositories that are not forks and have at least one pull request in the history. This filtering step ensures that the selected repositories demonstrate development activity, rather than being solely for archival purposes or mirrors of other open-source platforms such as BitBucket or GitLab~\cite{kalliamvakou2014promises}. Further refinement narrows down the selection to 886,365 by considering repositories that use Python or Java as their primary programming language. As the aim of this study is to investigate modifications within the README file, we focus only on repositories with at least 500 commits~\cite{wen2019large} to obtain repositories that went through code changes in different phases of the software development lifecycle. In this way, the README updates ideally capture a broader range of changes. Sampling from the entire population would give higher weight to repositories with fewer commits, potentially reducing diversity in the types of change reflected in README updates. We also used 10 stars as a threshold for eliminating toy projects~\cite{dabic2021sampling}. This further reduces the number of repositories to 28,896.

The earlier keyword filtering was applied to all commit messages without considering the semantics of changes specifically made to README files. Therefore, we further applied a filter on the README-changing commits, requiring them to have at least one occurrence in the keyword set, which decreased the dataset size to 14,958 records.

There are caveats that need to be taken care of when mining software repositories on GH Archive. First, some repositories are no longer available, either because they have been deleted or made private by their owners. After removing these non-available repositories, we retain 14,485 repositories. Second, due to renaming and other factors, some repositories are redirected by the GitHub API to alternative names, potentially causing duplicate repository issues. We further eliminated these repositories, reducing the number of repositories to 13,245. 

\begin{figure}[t!]
    \centering
    \includegraphics[width=\columnwidth]{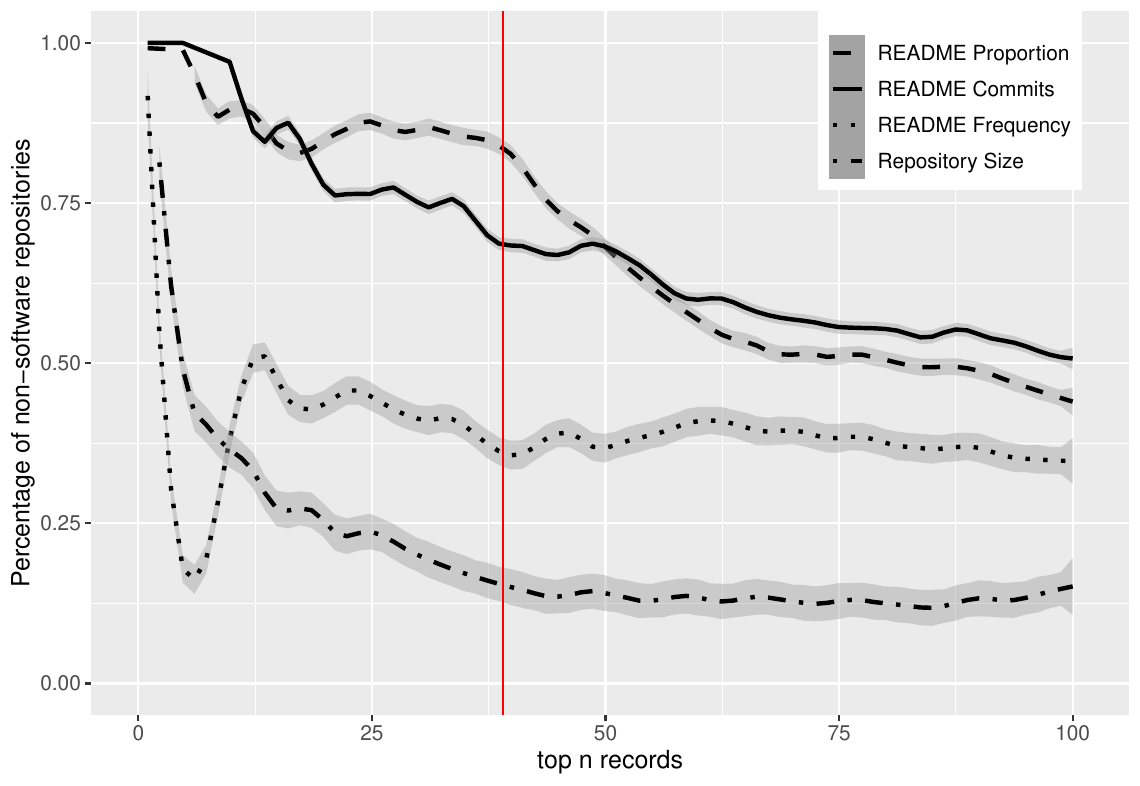}
    \caption{Ratio of non-software repositories among top n repositories}
    \label{fig:percentage_accumulated}
\end{figure}

In the last step, we removed outliers that fall outside of the scope of this study. We used FastText~\cite{joulin2016fasttext} to filter out 292 repositories with non-English README content. We manually inspected the repositories included after these filtering steps and found that there were some non-software repositories in there. Non-software repositories refer to projects primarily containing content other than traditional software code. These repositories may include, but are not limited to, collections of awesome lists, educational materials, and personal websites.  To exclude those systematically, we considered several factors that might indicate non-software projects, including (1) README commits to overall commits ratio, (2) number of README commits, (3) number of README commits per day, and (4) repository size. We sorted the repositories by these factors in descending order and manually labelled the top 100 records for each factor, respectively. Figure~\ref{fig:percentage_accumulated} illustrates how the percentage of non-software repositories varies as more projects are included. Although the number of README commits can identify the most unrelated repositories, the accuracy is merely 51\% for the top 100 entries. Meanwhile, regarding README commits to overall commit ratio, a plateau on the left side of the red vertical line indicates a consistently high accuracy of identifying non-software projects, with the red line representing the percentage of 52.9\% of this ratio. Therefore, we rounded the ratio to 50\%, used it as a threshold, and discarded these repositories, eventually obtaining a dataset consisting of 12,908 repositories.

\subsection{README Commit Classification}

We randomly sampled 400 repositories from the entire population, ensuring a confidence level above 95\% with a 5\% margin of error. For the 400 repositories sampled, there are 25,209 README commits in total, which need to be classified as ``relevant'' or ``irrelevant'' to the installation-related instructions. Given the prohibitively large number for manual annotation, we employed a semi-automated heuristic to significantly reduce the manual workload for identifying the relevant README updates.

Specifically, we extracted the different scales of headers from the sampled README files in their commit histories, with scales ranging from $\langle h1 \rangle$ to $\langle h6 \rangle$. We considered the occurrence of a particular header as one if it appeared at least once in the commit history. All repositories have at least one header, and Table~\ref{tab:header-scale-count} shows the distribution of header scales. Almost every repository has level 1 and level 2 headers, while level 3 and level 4 headers are also quite common. In contrast, the frequency of level 5 and level 6 headers is scarce. A closer examination identifies that level 5 and 6 do not convey meaningful document structural information; instead, some of them are used merely for emphasis within the text.


\begin{table}[h]
    \centering
    \caption{Counts for Header Scales}
    \begin{tabular}{c c c c c c c}
    \toprule
        Header Scale &  $\langle h1 \rangle$ & $\langle h2 \rangle$ & $\langle h3 \rangle$ & $\langle h4 \rangle$ & $\langle h5 \rangle$ & $\langle h6 \rangle$ \\
        \midrule
        Count  & 387 & 382 & 312 & 167 & 45 & 8\\
    \bottomrule
    \end{tabular}
    \label{tab:header-scale-count}
\end{table}

The basic idea of our heuristics is that README headers work as an overview of the contents beneath them. By examining the section headers, we can discern whether the sections are relevant to our target without delving into the detailed content. Therefore, we extracted the 100 most frequently occurring headers from $\langle h1 \rangle$ to $\langle h4 \rangle$, along with the 100 most frequently occurring word tokens. Three authors independently annotated these headers and tokens into categories of ``relevant'', ``irrelevant'' or ``not sure''. Agreement on the annotations was achieved through discussions in regular meetings. In total, we identified 66 relevant headers, 14 relevant keywords, and 177 irrelevant headers. These keywords and headers are also made available in our replication package.

Based on the information obtained from the headers, we developed a semi-automated labelling heuristic. Starting from the $\langle h1 \rangle$ headers, this program recursively examines whether the modified sections in the README commit are in the ``relevant'' set or not. If any of the modified sections has a ``relevant'' header, the program will mark the commit as relevant. Conversely, if all modified sections are categorised under the ``irrelevant'' headers, the commit is marked accordingly as irrelevant. Otherwise, the program cannot determine and proceeds to the next level of headers, repeating the decision process. If it still cannot ascertain the relevance after inspecting the headers at level 4, it assigns a ``not sure'' label, indicating that human annotation is required. 

To verify the accuracy of this program, we manually annotated four repositories, and the program reached an accuracy of 92.4\% (133/144). Subsequently, we applied this program to all repositories, resulting in 7,161 relevant, 8,583 irrelevant and 9,465 ``not sure'' commits generated by the semi-automatic labelling heuristic.

To annotate the data points categorised as ``not sure'', the first author randomly selected 50 records and three authors independently performed the annotation. The outcome of the annotation reached a Fleiss' Kappa score of 0.69, indicating a good level of agreement. The three authors further discussed the disagreed entries and came to an eventual agreement. Subsequently, the first author continued to annotate the remaining commits. Finally, within the dataset of 25,230 README commits, 8,673 commits were identified as relevant and 16,557 as irrelevant. The high accuracy rate (over 92.4\%) and high recall (82.6\%) indicate that our heuristic is quite capable of categorising the README documentation.

\subsection{Dataset Characteristics}
Before the details of the data analysis, we briefly introduce the characteristics of our curated dataset. The 8,673 identified relevant commits from the previous step cover 378 of 400 sampled repositories, that is, 94.5\% of the repositories contain at least one README commit modifying installation-related sections. To assess whether we had missed relevant installation-related documents, for each repository, we further searched ``\texttt{INSTALL.md}'', ``\texttt{INSTALLATION.md}'', ``\texttt{docs/INSTALL.md}'', and ``\texttt{docs/INSTALLATION.md}'' within its project folder (case insensitive), and consider them as additional installation documents other than the default README file. Only two repositories contain additional installation documents. Based on this observation, the majority of repositories in our analysed dataset use the README file as the primary installation documentation source, and the provided installation steps in the README file thus serve as the primary source for documentation users to access them. This observation further motivates us to focus on the analysis of the README file.

\begin{figure}
    \centering
    \includegraphics[width=\columnwidth]{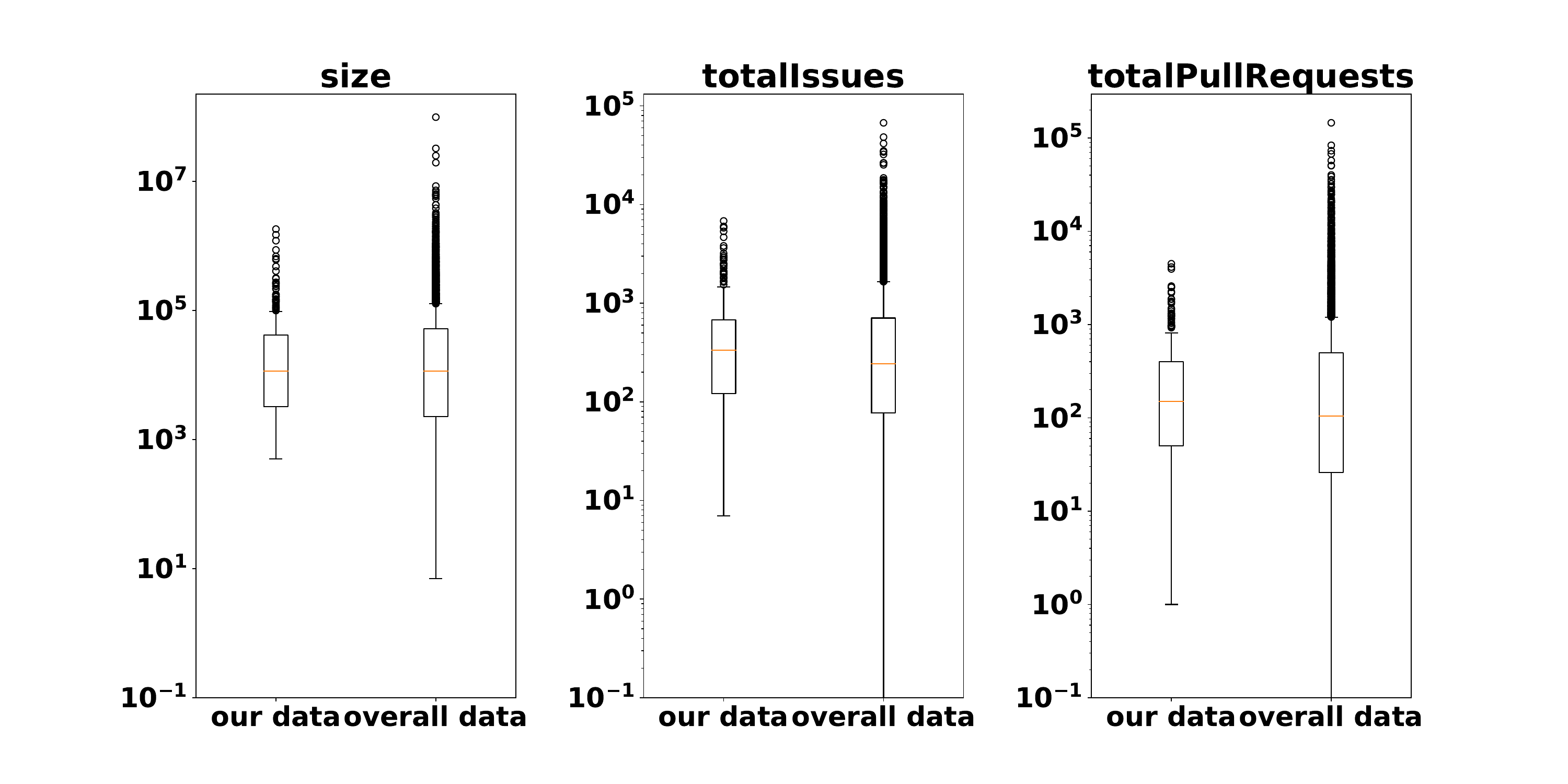}
    \caption{Comparison between repositories in our dataset and ones above 2,000 stars}
    \label{fig:size-compare-box}
\end{figure}

Regarding the representativeness of our analysed repositories, we selected three measures, namely repository size, total number of pull requests, and total number of issues in the repository. The repository size serves as a proxy for the complexity of the project, while the other measures quantify the project's interaction with its users. We accessed Java and Python repositories on Github with above 2,000 stars using GHS~\cite{dabic2021sampling}. The 2,000 stars threshold was also used in a previous study~\cite{chen2024large} when trying to build an Automatic Program Repair dataset, with the aim to collect data from popular repositories. Figure~\ref{fig:size-compare-box} shows the box plot for these three measures in our data and the repositories having above 2,000 stars. We performed the Mann-Whitney U test for these three measures, resulting in a p-value of 0.38, 0.02, and 0.06 for total size, total issues, and total pull requests, respectively. This indicates that there are no differences between the groups in the population in size and total pull requests. For total issues, we computed the Cohen's d effect size~\cite{becker2000effect}, which reached -0.09, indicating that the difference is negligible. Additionally, the majority of data from both sources ranges between 100 and 1000, suggesting a medium-to-large scale of issues in both sources. This result shows that our analysed projects are comparable to popular GitHub projects in terms of their scale and user interactions.

On average, each repository contains 23 relevant README commits, with a standard deviation of 25.7, indicating a skewed distribution. Specifically, with a minimum of 1 and a maximum of 224, the 25th, 50th, and 75th percentiles are 8, 14 and 31 commits, respectively. 

For the qualitative study, we need to obtain a sample from the data. However, due to the skewed distribution of README commits across repositories, a completely random sample would result in a dataset dominated by repositories with a large number of README commits, thereby reducing diversity across repositories.

Therefore, we performed a cluster sampling~\cite{baltes2022sampling}, splitting the dataset into four clusters: repositories with README commits from the minimum to the first quartile, from the first quartile to the second quartile, from the second quartile to the third quartile, and from the third quartile to the maximum. Within each cluster, we performed a random sample with the sample size required to achieve a 95\% confidence level (confidence interval: 5), resulting in samples of 206, 280, 324, and 358 README commits for each respective cluster. In total, we conducted a large-scale qualitative study on 1,168 README commits.


We calculated the correlation coefficient between the number of relevant README commits and the number of stars, as well as with the number of years the project has existed. The Pearson correlation coefficient reached 0.19 and 0.01, respectively. This statistic indicates that projects that actively participate in documentation activities, particularly for the installation-related sections, have a positive correlation to the project's popularity, while the number of relevant README commits does not necessarily grow with the years of the project. However, it is important to note that these correlations are relatively weak, while other factors also contribute to the popularity of repositories.

\begin{figure}[t!]
    \centering
    \includegraphics[width=\columnwidth]{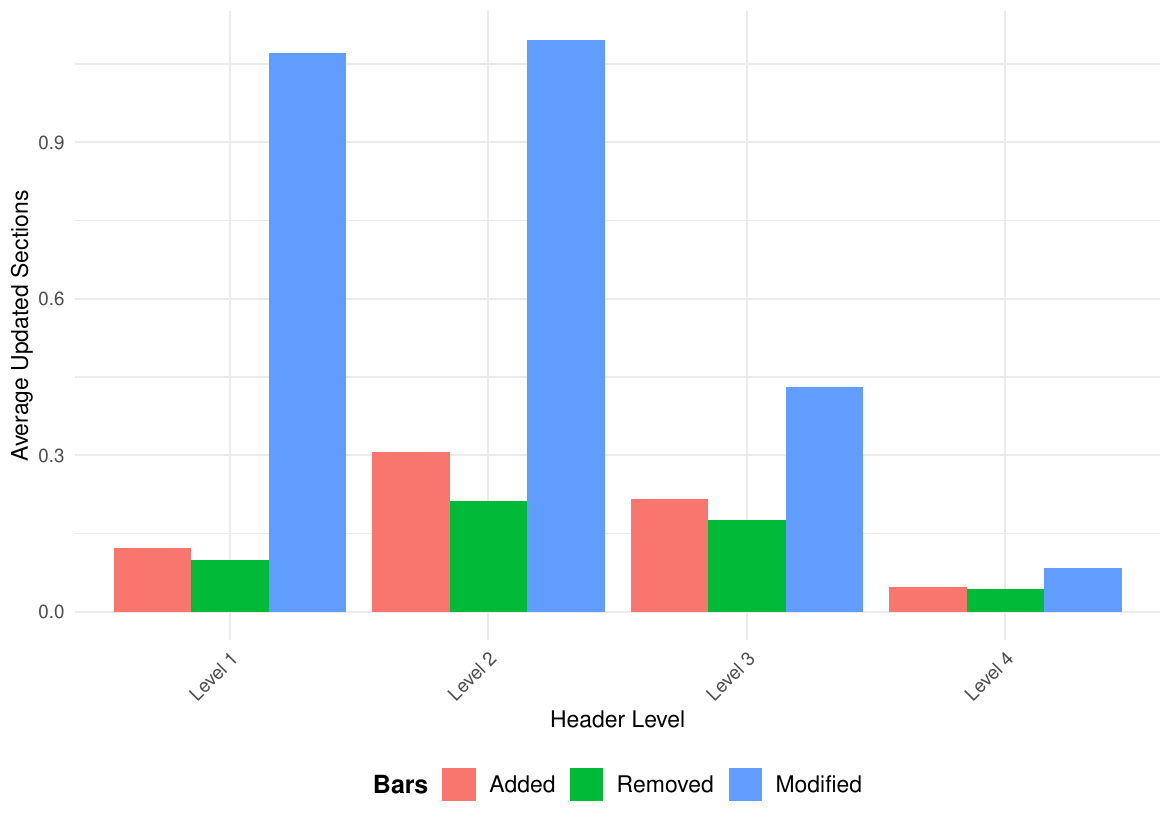}
    \caption{Average Updated Sections Per Commit}
    \label{fig:header-update-bar}
\end{figure}

From the section level, Figure~\ref{fig:header-update-bar} shows the number of sections that are added, removed, or modified per commit on average. The figure illustrates that the README updates mostly focus on sections in the top two hierarchies of headers, while the modification operation is much more frequent than the addition and removal operations.

\begin{figure}[t!]
     \centering
     \begin{subfigure}[b]{0.48\textwidth}
         \centering
         \includegraphics[width=\columnwidth]{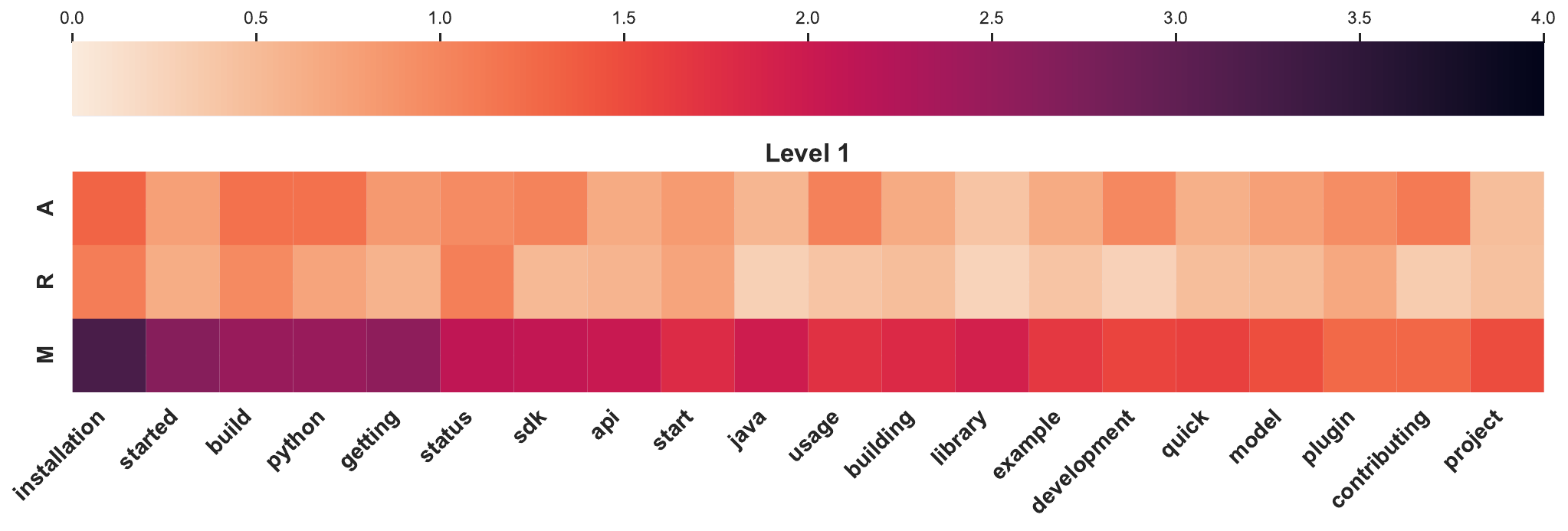}
         \caption{Level 1}
         \label{fig:level-1-heatmap}
     \end{subfigure}
     \hfill
     \begin{subfigure}[b]{0.48\textwidth}
         \centering
         \includegraphics[width=\columnwidth]{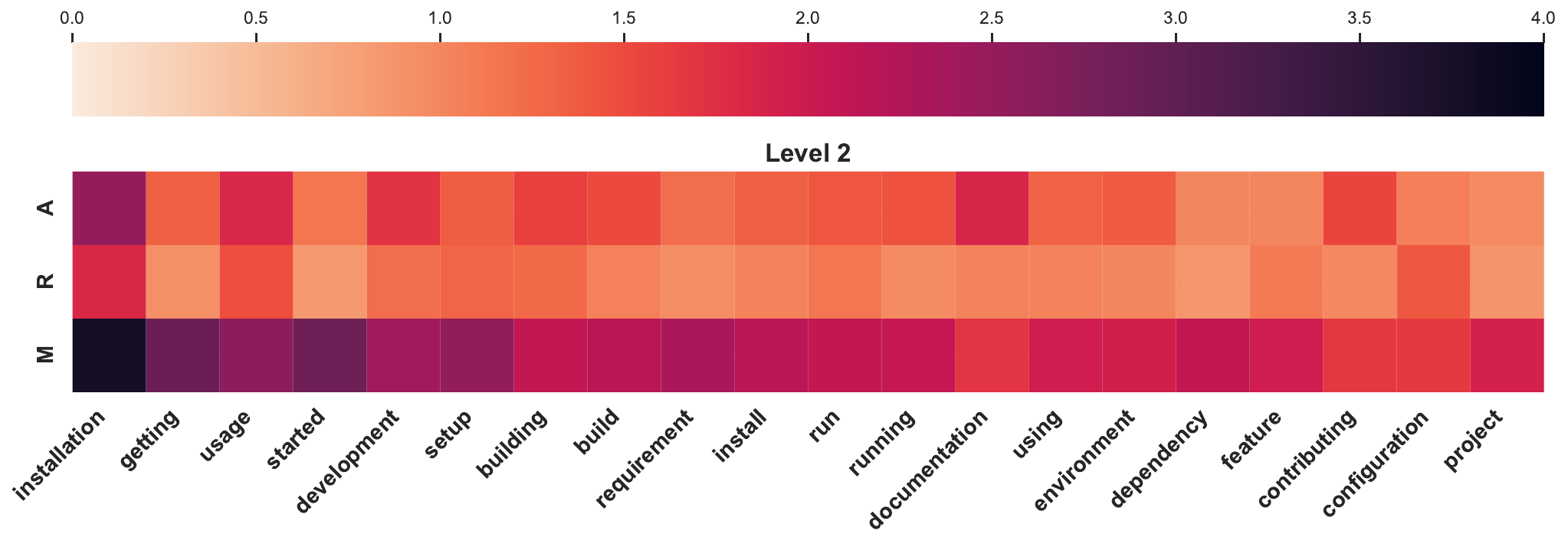}
         \caption{Level 2}
         \label{fig:level-2-heatmap}
     \end{subfigure}
        \caption{Update Keywords Heat-Map (A: Added, R: Removed, M: Modified)}
        \label{fig:keyword-heatmap}
\end{figure}

Based on the results of Figure~\ref{fig:header-update-bar}, we visualised the top 20 most updated keywords within the section headers at the first two levels to have a better understanding of the updated sections. The heat maps are presented in Figure~\ref{fig:keyword-heatmap}.

From the figures, it is clear that more addition and removal operations are conducted on the level 2 hierarchy, while most sections are related to themes including installation, usage, build, setup and running.

\subsection{Data Analysis}

We conducted a qualitative study of the annotated data to answer our research question. Specifically, we synthesised a comprehensive taxonomy of 6 categories, consisting of 189 different codes. The findings provide a comprehensive understanding of the nature of the modifications with rich implications for practitioners and researchers.

We adopted open coding~\cite{glaser2017discovery} for qualitatively analysing the update events. The first author conducted the data analysis and created spreadsheets containing all raw data and codes. The spreadsheets were shared among all authors, and the second and third authors cross-validated all codes, concepts, and categories to reduce bias and increase the reliability of the findings~\cite{service2009book}. Concepts and categories were discussed among authors through regular meetings and finalised after several rounds of revisions. 

\begin{figure*}[t!]

    \centering
    \begin{subfigure}[b]{1.9\columnwidth}
    \includegraphics[width=1\columnwidth]{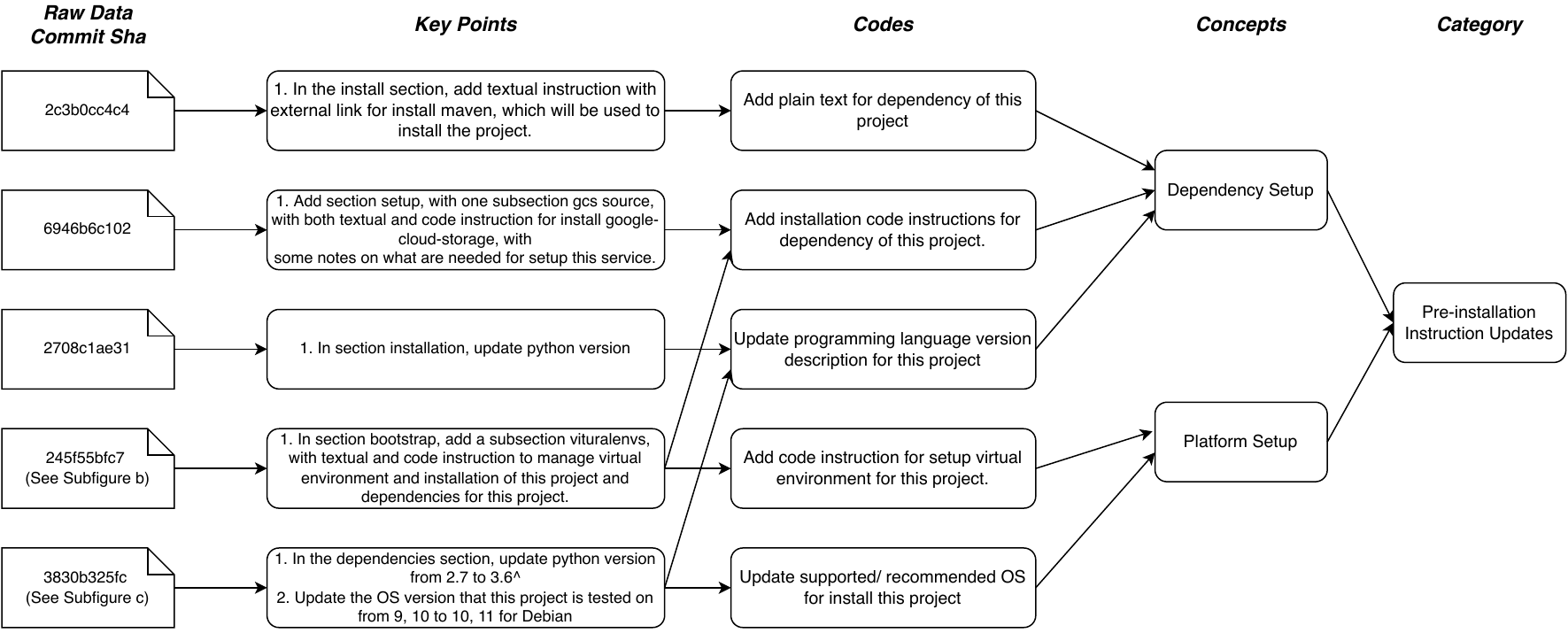}
    \caption{Steps for Synthesising Categories from Raw Data}
    \label{fig:coding-process}
     \end{subfigure}
    \vspace{0.5cm}
     
    \begin{subfigure}[b]{0.95\columnwidth}
    \centering
    \includegraphics[width=\columnwidth]{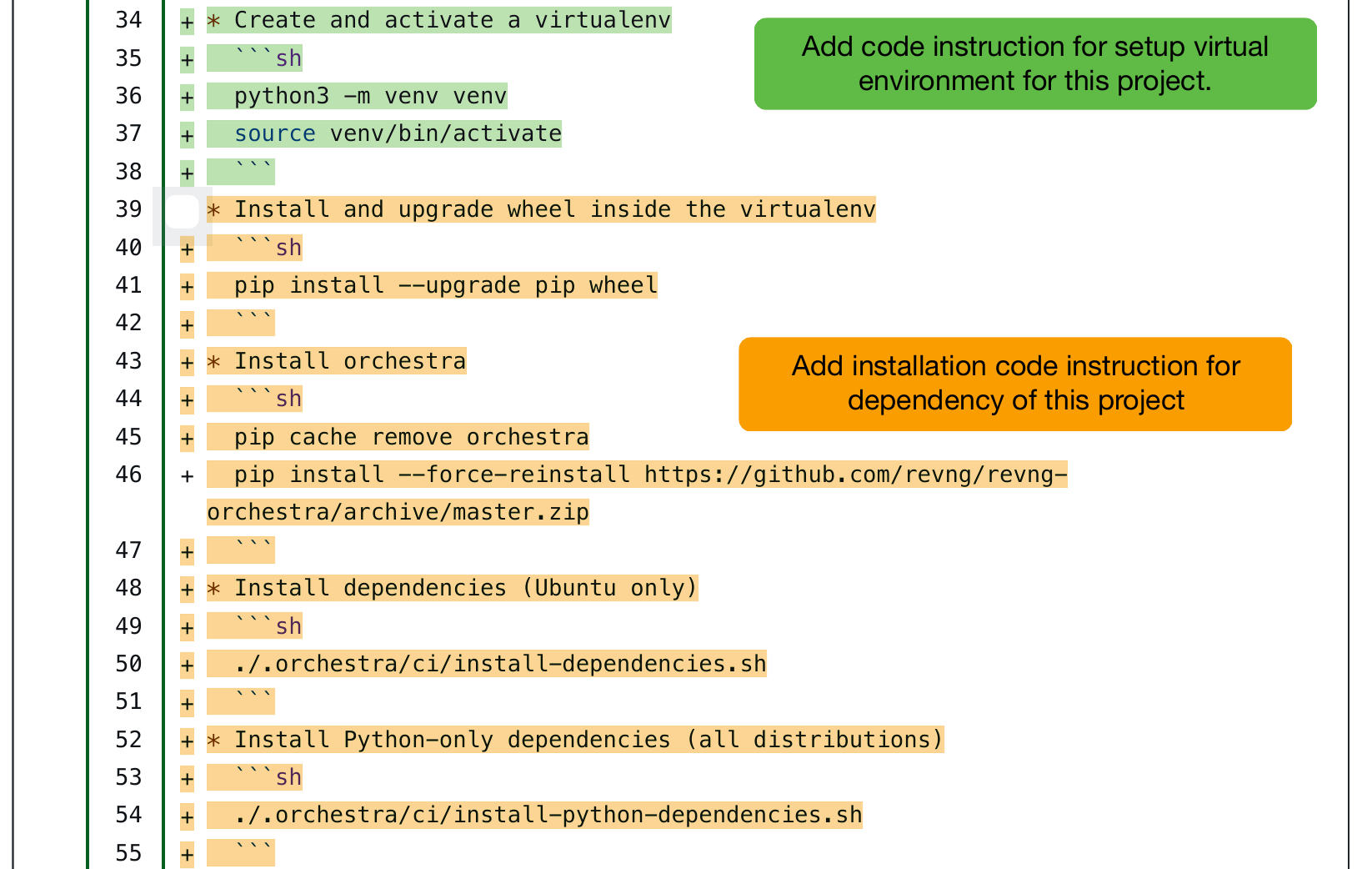}
    \caption{Example for Commit \texttt{245f55bfc7}}
    \label{fig:Process-Example-1}
  \end{subfigure}%
  \begin{subfigure}[b]{0.95\columnwidth}
    \centering
    \includegraphics[width=\columnwidth]{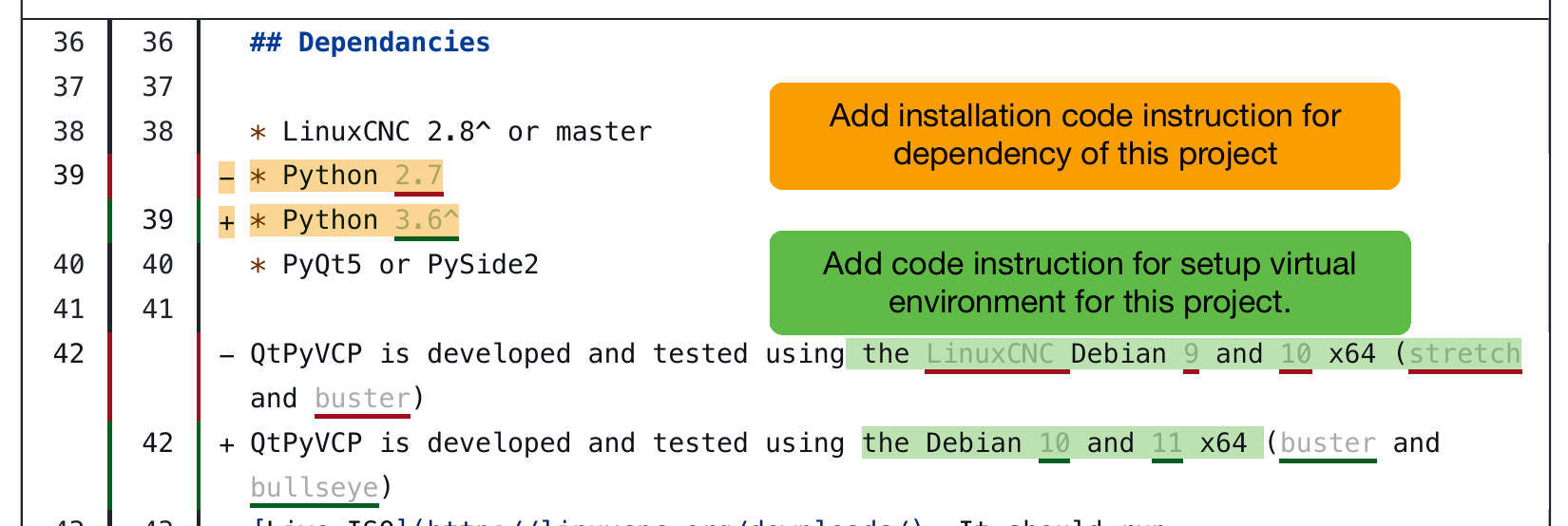}
    \caption{Example for Commit \texttt{3830b325fc}}
    \label{fig:Process-Example-2}
  \end{subfigure}
  \caption{Coding Process and Examples of Assigned Codes}
  \label{fig:coding-process-overall}
     
\end{figure*}

We started our analysis by searching each commit on GitHub and recording the text summarising the changes from the perspective according to our research question: updates for the installation-related sections. We call this process ``key points extraction and getting familiar with the data''. Then we rigorously performed open coding to assign each README commit with one to many succinct descriptions of their update modifications. This process took into account both the key points extracted and the content of the README commits. The additional investigation into the raw README commit data is to address any unclear key points and to double-check whether any part of the modification is missing. 

During the coding procedure, we constantly compared the emerging codes within and across different README commits to synthesise a higher level of abstraction called concepts, and later the highest level of abstraction called categories. Figure~\ref{fig:coding-process} illustrates examples of the procedure of generating one of the categories called ``Pre-installation Instruction Updates''. Figures~\ref{fig:Process-Example-1} and Figure~\ref{fig:Process-Example-2} further explain the process for how we assign codes for different changes within the commit. We have made our analysed data available. Please refer to the Replication Package section for more details.

\begin{figure}[ht!]
    \centering
    \includegraphics[width=\columnwidth]{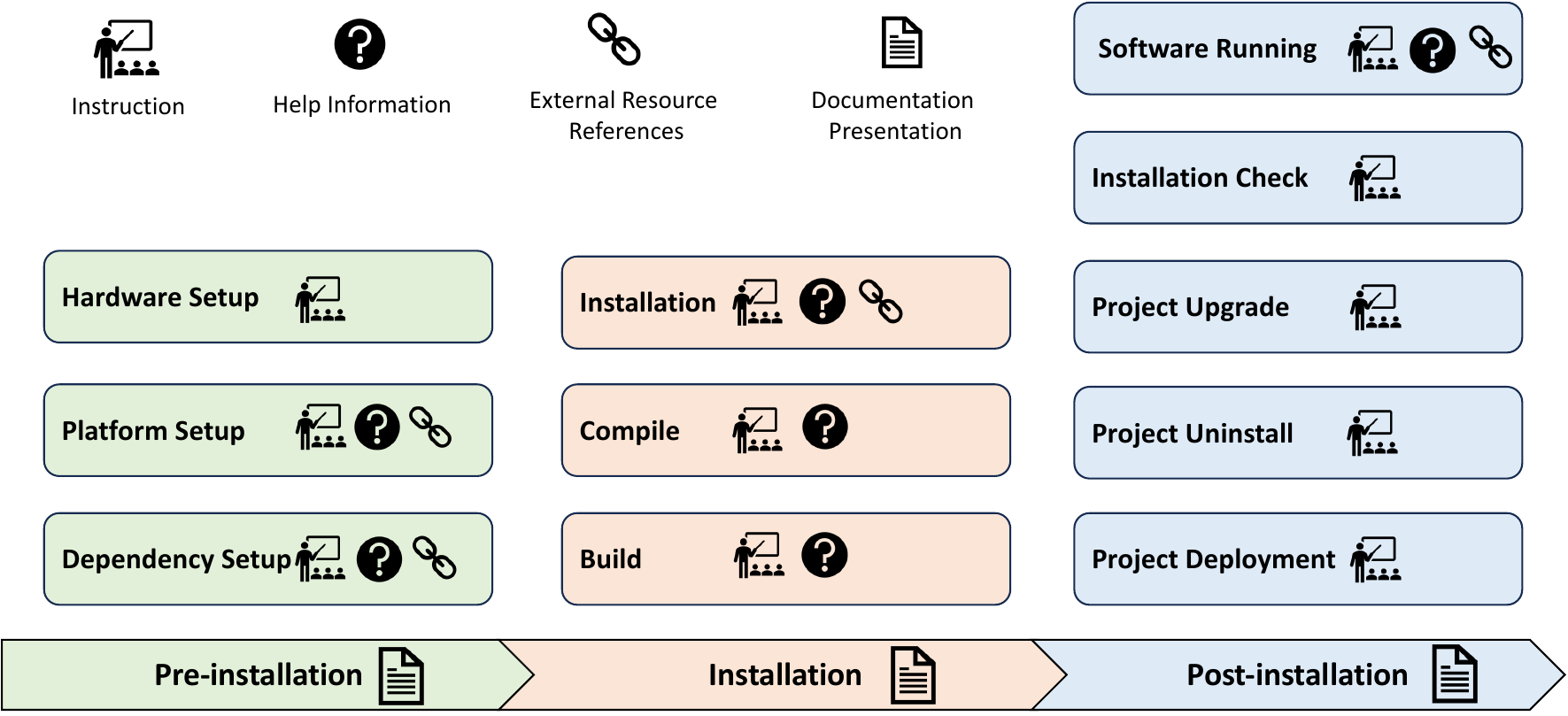}
    \caption{Development Tasks in README Files}
    \label{fig:development-task}
\end{figure}

\section{Results}
Table~\ref{tab:taxonomy} presents the hierarchical taxonomy of 189 codes that we identified from the 1,168 data points to answer our RQ1. The updates are categorised into six main groups: (1) \textit{pre-installation instruction}; (2) \textit{installation-related instruction}; (3) \textit{post-installation instruction}; (4) \textit{help information}; (5) \textit{document presentation}; and (6) \textit{external resource management}.

To give readers a better understanding of our analysed result, Figure~\ref{fig:development-task} synthesises our analysis in a timeline fashion, corresponding to pre-installation, installation, and post-installation, where different software development tasks are included in each phase. Moreover, various information perspectives regarding the README files are also displayed within this figure, including instructions, help information, external resource links, and document presentation. For each of the categories, we describe representative examples and discuss implications for practitioners and researchers.

It is worth noting that, compared to other qualitative studies~\cite{aghajani2019software, aghajani2020software, prana2019categorizing, uddin2015api}, our work focuses on modifications to Markdown-formatted text, which involves two pieces of documents. Due to the substantial space to illustrate examples of README updates, we will only include figures for a few examples. Additional examples are made accessible through URLs that we have formatted as clickable links in Table~\ref{tab:taxonomy}. We have also included the URLs and their corresponding screenshots in our replication package.

\begin{table*}
    \centering
    \caption{Examples for Qualitative Analysis}

    \begin{tabular}{p{2cm}p{4.9cm}rrlrrr}
        \toprule
        Category & Concept & \# Codes & Frequency  & Example ID & Addition & Modification & Removal \\
        \midrule
        \multirow{4}{*}{\shortstack[l]{Pre-installation \\ instruction}} 
        & Dependency Setup & 23 & 527 (45.3\%) & \href{https://github.com/karlch/vimiv-qt/commit/2d6d454bc78fa95fd0c9b810b53a349e36aa7490##diff-b335630551682c19a781afebcf4d07bf978fb1f8ac04c6bf87428ed5106870f5R10}{{\textcolor{blue}{1}}}, \href{https://github.com/chowdhary987/blueocean-pulgin/commit/3d14e8440634fd0820be84e2d5d2791e8dd60474##diff-b335630551682c19a781afebcf4d07bf978fb1f8ac04c6bf87428ed5106870f5R107}{\textcolor{blue}{2}}, \href{https://github.com/openmaraude/APITaxi/commit/38a18002b514d7c43ee230f36c6b7f705300f44e##diff-b335630551682c19a781afebcf4d07bf978fb1f8ac04c6bf87428ed5106870f5R32}{\textcolor{blue}{3}}, \href{https://github.com/nextstrain/cli/commit/1556e7484d9ed4077cf25520d317944f2c8ec36d##diff-b335630551682c19a781afebcf4d07bf978fb1f8ac04c6bf87428ed5106870f5R96}{\textcolor{blue}{4}} & 47.8\% & 34.8\% & 17.4\% \\
        & Platform Setup & 11 & 82 (7.1\%) & \href{https://github.com/revng/orchestra/commit/245f55bfc7092868440954d1f877b0cff0a42996##diff-b335630551682c19a781afebcf4d07bf978fb1f8ac04c6bf87428ed5106870f5R32}{\textcolor{blue}{5}}, \href{https://github.com/rizac/stream2segment/commit/181b95c5e472d2dd45c8120ecc07e927534593b1##diff-b335630551682c19a781afebcf4d07bf978fb1f8ac04c6bf87428ed5106870f5R185}{\textcolor{blue}{6}}, \href{https://github.com/kcjengr/qtpyvcp/commit/3830b325fcbf767e4104d7402acac1a7f82c1929##diff-b335630551682c19a781afebcf4d07bf978fb1f8ac04c6bf87428ed5106870f5R42}{\textcolor{blue}{7}} & 54.5\% & 18.2\% & 27.3\%\\
        & Virtual Machine and Hardware & 6 & 19 (1.6\%) & ------ & 66.7\% & 16.7\% & 16.7\% \\
        \cmidrule(l){2-8}
        & \textbf{Total} & \textbf{40} & \textbf{628 (54.0\%)} & & \textbf{52.5\%} & \textbf{27.5\%} & \textbf{20.0\%} \\
        \midrule
        \multirow{3}{*}{\shortstack[l]{Installation-related \\ instruction}} 
        & Source code & 23 & 267 (23.0\%) & \href{https://github.com/SUNCAT-Center/CatLearn/commit/45e8e6cd1e5475db85174b5fc10299e16dcf69d0##diff-b335630551682c19a781afebcf4d07bf978fb1f8ac04c6bf87428ed5106870f5R50}{\textcolor{blue}{8}}, \href{https://github.com/akto-api-security/akto/commit/4ef6d2fb3f9b0d448c1b333df92589bc6e973635##diff-b335630551682c19a781afebcf4d07bf978fb1f8ac04c6bf87428ed5106870f5R45}{\textcolor{blue}{9}} & 43.4\%& 43.5\% & 13.0\% \\
        & Published package & 14 & 162 (13.9\%) & \href{https://github.com/mneyapo/flexx/commit/7f79092c9cf7784ed05d02a3e5a0529888d9be51##diff-b335630551682c19a781afebcf4d07bf978fb1f8ac04c6bf87428ed5106870f5R62}{\textcolor{blue}{10}}, \href{https://github.com/ashokoripella/ashokoripella/commit/e63348db5cfcfeaf5e8e9b40d04bdeb067a5008a##diff-b335630551682c19a781afebcf4d07bf978fb1f8ac04c6bf87428ed5106870f5R71}{\textcolor{blue}{11}}, \href{https://github.com/ut-parla/Parla.py/commit/ed056c29bc18159a008208f3c5456f7a742acdc8##diff-b335630551682c19a781afebcf4d07bf978fb1f8ac04c6bf87428ed5106870f5R59}{\textcolor{blue}{12}}, \href{https://github.com/osmdroid/osmdroid/commit/d4dde39a02e37c93e5552b540949d8a4a3543b41##diff-b335630551682c19a781afebcf4d07bf978fb1f8ac04c6bf87428ed5106870f5R27}{\textcolor{blue}{13}} & 57.1\% & 28.6\% & 14.3\%\\
        \cmidrule(l){2-8}
        & \textbf{Total} & \textbf{37}& \textbf{429 (36.9\%)} & & \textbf{48.6\%} & \textbf{37.8\%} & \textbf{13.5\%} \\
        \midrule
        \multirow{7}{*}{\shortstack[l]{Post-installation \\ instruction}} 
        & Project running instruction & 12 & 185 (15.9\%) & \href{https://github.com/isl-org/Open3D-ML/commit/32cd56206147f132341464df6d39a70eecb78397##diff-b335630551682c19a781afebcf4d07bf978fb1f8ac04c6bf87428ed5106870f5R135}{\textcolor{blue}{14}}, \href{https://github.com/uds-se/FormatFuzzer/commit/d257145ca1bd6db23f1621ed18b8e72a14057f37##diff-b335630551682c19a781afebcf4d07bf978fb1f8ac04c6bf87428ed5106870f5R45}{\textcolor{blue}{15}} & 37.5\% & 50.0\% & 12.5\% \\
        & Installation check & 3 & 13 (1.1\%) & \href{https://github.com/genixpro/kwola/commit/f5e63fcef8c60330d0c101ed87efedcbfc0b8c22##diff-b335630551682c19a781afebcf4d07bf978fb1f8ac04c6bf87428ed5106870f5R166}{\textcolor{blue}{16}} & 40.0\% & 60.0\% & 0.0\% \\
        & Running Tests & 6 & 32 (2.8\%) & ------ & 88.9\% & 0.0\% & 11.1\% \\
        & Deployment Instruction & 3 & 9 (0.8\%) & ------ & 75\% & 0.0\% & 25.0\% \\
        & Upgrade Instruction & 4 & 6 (0.5\%) & ------ & 100.0\% & 0.0\% & 0.0\% \\
        & Uninstall Instruction & 2 & 3 (0.3\%) & ------ & 0.0\% & 0.0\% & 100.0\% \\
        \cmidrule(l){2-8}
        & \textbf{Total} & \textbf{30} & \textbf{248 (21.3\%)} & & \textbf{58.6\%} & \textbf{24.1\%} & \textbf{17.2\%} \\
        \midrule
        \multirow{4}{*}{\shortstack[l]{Help information}} 
        & Tutorial updates & 6 & 70 (6.0\%) & \href{https://github.com/MCXA/Phenotyping/commit/17d922999851c8c8f1e368be500367187ca33f2c##diff-b335630551682c19a781afebcf4d07bf978fb1f8ac04c6bf87428ed5106870f5R38}{\textcolor{blue}{17}} & 50.0\% & 33.3\% & 16.7\% \\
        & Explanation & 14 & 125 (10.7\%) & \href{https://github.com/SUNCAT-Center/CatLearn/commit/01bf3e9d634fdb19372756a1e7f7403aa178d667##diff-b335630551682c19a781afebcf4d07bf978fb1f8ac04c6bf87428ed5106870f5R67}{\textcolor{blue}{18}}, \href{https://github.com/SUNCAT-Center/CatLearn/commit/1e68b55808c37e2698bc082556dd7e6160bf2435##diff-b335630551682c19a781afebcf4d07bf978fb1f8ac04c6bf87428ed5106870f5R46}{\textcolor{blue}{19}} & 61.5\% & 38.5\% & 0.0\% \\
        & Notes and troubleshooting & 16 & 70 (6.0\%) & \href{https://github.com/awslabs/aws-crt-java/commit/dad3837984c3bfe4e260dedd4196a95dc9b88d17##diff-b335630551682c19a781afebcf4d07bf978fb1f8ac04c6bf87428ed5106870f5R99}{\textcolor{blue}{20}}, \href{https://github.com/decentralized-identity/universal-resolver/commit/81553bd600bf31cf61a3954c16fe54a4e3f33e20##diff-b335630551682c19a781afebcf4d07bf978fb1f8ac04c6bf87428ed5106870f5R51}{\textcolor{blue}{21}} & 68.8\% & 6.3\% & 25.0\% \\
        \cmidrule(l){2-8}
        & \textbf{Total} & \textbf{36} & \textbf{265 (22.8\%)} & & \textbf{62.9\%} & \textbf{22.9\%} & \textbf{14.3\%}\\
        \midrule
        \multirow{6}{*}{\shortstack[l]{Document \\ Presentation}} 
        & Documentation structural modification & 7 & 197 (16.9\%) & \href{https://github.com/itsallcode/white-rabbit/commit/bd1cefc49d728b8bbcbdabc0df169d9a44491784##diff-b335630551682c19a781afebcf4d07bf978fb1f8ac04c6bf87428ed5106870f5}{\textcolor{blue}{22}}, \href{https://github.com/actinia-org/actinia-core/commit/4f7363b74033a40b3265e1cfc970bb2ee59bea55##diff-b335630551682c19a781afebcf4d07bf978fb1f8ac04c6bf87428ed5106870f5R47}{\textcolor{blue}{23}} & 0.0\% & 100.0\% & 0.0\% \\
        & Content formatting and cleaning & 10 & 119 (10.2\%) & \href{https://github.com/jupyterhub/dockerspawner/commit/4a90a55e8449d3d303e96610e41b2ef137b51811##diff-b335630551682c19a781afebcf4d07bf978fb1f8ac04c6bf87428ed5106870f5R69}{\textcolor{blue}{24}}, \href{https://github.com/Tolu-gith/tt/commit/907705873708227d568a72d5e3b200713c6b14d4##diff-b335630551682c19a781afebcf4d07bf978fb1f8ac04c6bf87428ed5106870f5R24}{\textcolor{blue}{25}}, \href{https://github.com/alorence/django-modern-rpc/commit/d2d028aae74e328c59aae06e8142c2b67cb1d54a##diff-b335630551682c19a781afebcf4d07bf978fb1f8ac04c6bf87428ed5106870f5R36}{\textcolor{blue}{26}}, \href{https://github.com/darkwizard242/cis_ubuntu_2004/commit/c26a1c99a1427d1cfe0985b7443a27f0d99ce918##diff-b335630551682c19a781afebcf4d07bf978fb1f8ac04c6bf87428ed5106870f5R379}{\textcolor{blue}{27}} & 0.0\% & 100.0\% & 0.0\% \\
        & Presentation fixing & 2 & 61 (5.2\%) & \href{https://github.com/PMCC-BioinformaticsCore/janis-core/commit/c72096c153e94f4c65709e02ee7c36606cccf61b##diff-b335630551682c19a781afebcf4d07bf978fb1f8ac04c6bf87428ed5106870f5R6}{\textcolor{blue}{28}}, \href{https://github.com/MaximProkhorov/vera-pdf-library/commit/1a86148b0dc154eb14c2dfd14a07fe92a1ee747f##diff-0324feccee544c74b7e269abb8ef3840e485a44db47736ad2ac3d8d658b415edR9}{\textcolor{blue}{29}} & 0.0\% & 100.0\% & 0.0\%\\
        & Text rephrasing and editing & 5 & 70 (6.0\%) & \href{https://github.com/actris-cloudnet/cloudnetpy/commit/458623df6af9bd84451f291b2dff256120d2c265##diff-b335630551682c19a781afebcf4d07bf978fb1f8ac04c6bf87428ed5106870f5R50}{\textcolor{blue}{30}}, \href{https://github.com/Ravaelles/Atlantis/commit/89fe47619edf026500c1db491679bd4fd3b8d7f7##diff-b335630551682c19a781afebcf4d07bf978fb1f8ac04c6bf87428ed5106870f5R25}{\textcolor{blue}{31}} & 0.0\% & 100.0\% & 0.0\% \\
        & Documentation debt & 4 & 30 (2.6\%) & \href{https://github.com/Kuifje02/vrpy/commit/bd6b882db0fe99f6a78570d94312fe6277293061##diff-b335630551682c19a781afebcf4d07bf978fb1f8ac04c6bf87428ed5106870f5R43}{\textcolor{blue}{32}} & 25.0\% & 0.0\% & 75.0\% \\
        \cmidrule(l){2-8}
        & \textbf{Total} & \textbf{28} & \textbf{477 (41.0\%)} & & \textbf{3.6\%} & \textbf{85.7\%} & \textbf{10.7\%}  \\
        \midrule
        \multirow{4}{*}{\shortstack[l]{External resources \\ management}} 
        & Project external documentation & 3 & 120 (10.3\%) & \href{https://github.com/agonyforge/arbitrader/commit/ee31e68e547e747dafbfd443e7718ebc373c73b8##diff-b335630551682c19a781afebcf4d07bf978fb1f8ac04c6bf87428ed5106870f5R12}{\textcolor{blue}{33}} & 25.0\% & 50.0\% & 25.0\%  \\
        & Project non-documentation artefact & 11 & 136 (11.7\%) & \href{https://github.com/sdauzcm/SR-basicSR/commit/de3ba707e538aa1d8395376a9fae5e20c9c4ef40##diff-b335630551682c19a781afebcf4d07bf978fb1f8ac04c6bf87428ed5106870f5R44}{\textcolor{blue}{34}} & 44.4\% & 33.3\% & 22.2\% \\
        & Third-party external resources & 4 & 110 (9.5\%) & \href{https://github.com/BlueBrain/NeuroMorphoVis/commit/f32210d72a4ac9141b3f5b5af6e8231d0de7ef4c##diff-b335630551682c19a781afebcf4d07bf978fb1f8ac04c6bf87428ed5106870f5R13}{\textcolor{blue}{35}} & 60.0\% & 20.0\% & 20.0\% \\
        \cmidrule(l){2-8}
        & \textbf{Total} & \textbf{18} & \textbf{366 (31.5\%)} & & \textbf{44.4\%} & \textbf{33.3\%} & \textbf{22.2\%} \\        
        \bottomrule
    \end{tabular}
    \label{tab:taxonomy}
\end{table*}

\subsection{Pre-installation Instruction}
Pre-installation instructions, also seen as environment setup instructions, contain preparation instructions before installation. In our analysed samples, 628 out of 1,168 performed updates are related to environment setup topics in their README documents. In this category, we synthesised three concepts, which are discussed as follows:

\textbf{Dependency Setup (527)}.\footnote{Number in the parenthesis represents the frequency of this concept.} Dependencies enable code reuse and require maintenance effort as the project and other dependencies evolve~\cite{bogart2021and}. In our analysed dataset, more than 40\% of the documentation updates contain dependency-related instructional modifications, indicating the importance of this perspective.

\begin{figure}[ht!]
    \centering
    \begin{subfigure}[b]{\columnwidth}
         \centering
         \includegraphics[width=\columnwidth]{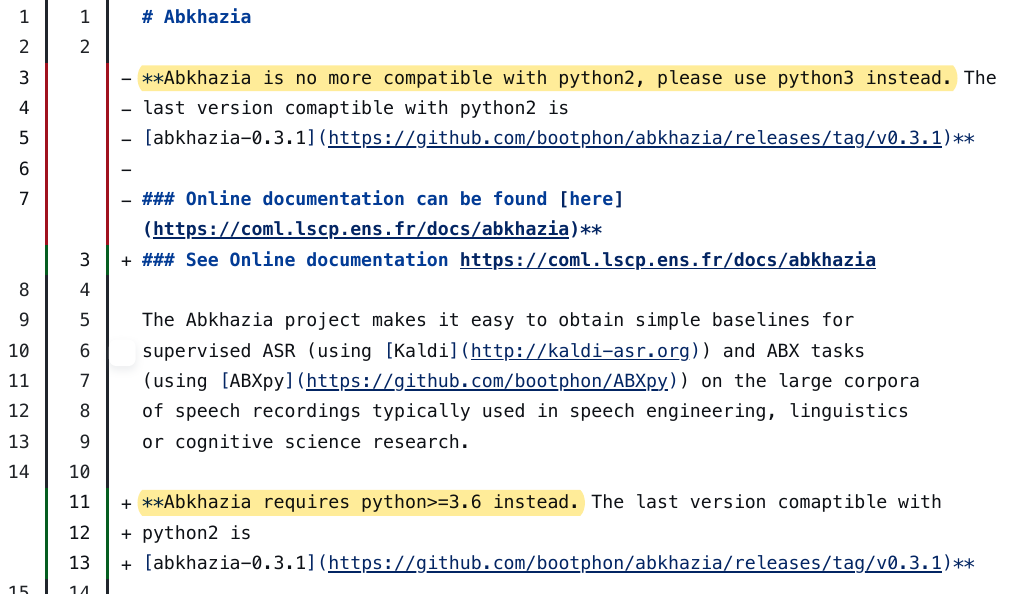}
         \caption{}
         \label{fig:Programming-Lnaguage-Example-Update1}
     \end{subfigure}

     \begin{subfigure}[b]{\columnwidth}
         \centering
         \includegraphics[width=\columnwidth]{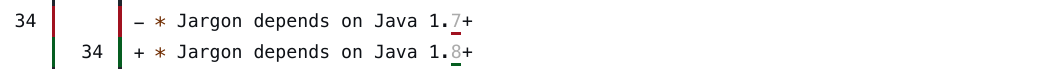}
         \caption{}
         \label{fig:Programming-Lnaguage-Example-Update2}
     \end{subfigure}

      \begin{subfigure}[b]{\columnwidth}
         \centering
         \includegraphics[width=\columnwidth]{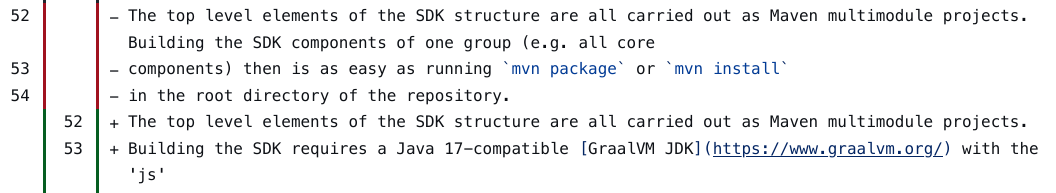}
         \caption{}
         \label{fig:Programming-Lnaguage-Example-Update3}
     \end{subfigure}

     \caption{Programming Language Examples}
     \label{fig:Programming-Language-Examples}
\end{figure}

Among the modifications of the dependency setup, the concept of programming language version compatibility refers to details concerning the target version of the programming language upon which the software is built, ensuring its accurate execution. Figure~\ref{fig:Programming-Language-Examples} includes examples of how the programming language version information in the README files is updated. Updates are mostly small, only reflecting the new version of the programming language that their projects are using. However, as seen in Figure~\ref{fig:Programming-Lnaguage-Example-Update3}, when the programming language requirement information is newly added to the documentation, it means that the previous README suffered from incomplete information, and efforts are made to remedy the lack of information in the programming language requirements.

Apart from the programming language, other dependencies setup include external libraries and modules (e.g., API), as well as third-party software (e.g., database) that projects rely on. The inclusion of new features, the resolution of technical debt during project development, and the updates of constantly evolving dependencies collectively contribute to the volatility of the dependency management information within the README files. Indeed, most updates in this aspect involve updating dependency information, as indicated in \href{https://github.com/karlch/vimiv-qt/commit/2d6d454bc78fa95fd0c9b810b53a349e36aa7490#diff-b335630551682c19a781afebcf4d07bf978fb1f8ac04c6bf87428ed5106870f5R10}{\textit{Example 1}}, and instructions on installation of dependencies as indicated in \href{https://github.com/chowdhary987/blueocean-pulgin/commit/3d14e8440634fd0820be84e2d5d2791e8dd60474#diff-b335630551682c19a781afebcf4d07bf978fb1f8ac04c6bf87428ed5106870f5R107}{\textit{Example 2}}. We observed a common practice of manifesting dependencies by listing them (e.g., in the forms of plain texts, bullet points, or tables), with additions, updates, and deletions to reflect the corresponding changes.  Meanwhile, the patterns for providing the installation instruction of dependencies are quite diverse. Centralised commands (e.g., \texttt{requirements.txt}) or separate installation commands for each dependency are widely used. However, maintaining the separate command style requires more effort, as individual attention to each dependency is needed, making modifications potentially more time-consuming and the process more complicated.

In \href{https://github.com/openmaraude/APITaxi/commit/38a18002b514d7c43ee230f36c6b7f705300f44e#diff-b335630551682c19a781afebcf4d07bf978fb1f8ac04c6bf87428ed5106870f5R32}{\textit{Example 3}}, the commit adds configuration instructions for the database on which the project relies. This underscores the importance of incorporating essential configuration details when dependencies are not ready for use following the dependency installation step, which is a common scenario for third-party software dependencies. Interestingly, some examples, such as commit \href{https://github.com/nextstrain/cli/commit/1556e7484d9ed4077cf25520d317944f2c8ec36d#diff-b335630551682c19a781afebcf4d07bf978fb1f8ac04c6bf87428ed5106870f5R96}{\textit{Example 4}}, add guidelines to verify the success of dependency installation.

\textbf{Platform Setup (82)}. Platform setup information includes details for systems on which the software operates, which is the second largest updated concept for the pre-installation instructions category. 


In our analysed data, we found that 48 instances involve modification in setting up virtual environments, and the majority of them are related to newly added instructions such as the one indicated in \href{https://github.com/revng/orchestra/commit/245f55bfc7092868440954d1f877b0cff0a42996#diff-b335630551682c19a781afebcf4d07bf978fb1f8ac04c6bf87428ed5106870f5R32}{\textit{Example 5}}. Furthermore, \href{https://github.com/rizac/stream2segment/commit/181b95c5e472d2dd45c8120ecc07e927534593b1#diff-b335630551682c19a781afebcf4d07bf978fb1f8ac04c6bf87428ed5106870f5R185}{\textit{Example 6}} explicitly mentions the virtual environment as their recommended way of installation of the project.

Operating system information is another platform setup concern. We observed efforts that mostly focused on adding operating system compatibility information (e.g., \textit{BiglyBT comes in several editions for different operating systems. Mac OSX, Linux and Windows use the full BiglyBT-API based on Java 8.}) or updating operating system version information as indicated in \href{https://github.com/kcjengr/qtpyvcp/commit/3830b325fcbf767e4104d7402acac1a7f82c1929#diff-b335630551682c19a781afebcf4d07bf978fb1f8ac04c6bf87428ed5106870f5R42}{\textit{Example 7}}, which updated the support operating system from Debian 9 and 10 to Debian 10 and 11.

\textbf{Virtual Machine and Hardware (19)}. Depending on the nature and usage purpose of different software repositories, hardware requirements may differ for different software projects and sometimes even require deployment on cloud servers. We observed commit updates that include hardware specification information or steps for configuring a virtual machine. It would be important to include this information for certain projects, although this does not generalise to all repositories.

\begin{tcolorbox}[left=1pt, top=1pt, right=1pt, bottom=1pt]
    \textbf{Summary of pre-installation instruction results}: The concept of dependency setup includes management of programming language version and other dependencies to reflect software system changes. However, the concept of platform setup majorly includes updates in providing virtual environment instructions and operating system information of the software project. Virtual machine and hardware instructions are less frequently encountered.
\end{tcolorbox}
 

\subsection{Installation-related Instruction}

In this category, we synthesised two concepts, which are discussed as follows:

\textbf{Installation-related Instructions from Source Code Operations (267)}. Installation from source code requires users to obtain the source code and then locally execute a set of predefined operations, encompassing compilation, building, and other procedures, to install the software project. While projects may use various technologies, including Docker, Pip, and Maven, we identified in our analysed data that source code-level installation is the most commonly adopted method that the projects tend to initially provide. In addition, the inclusion of newly added instructions also accounts for various operating systems. Given that commands for Windows systems typically differ from those for Linux, additional information in the README file instructions or even the source code is needed to facilitate a smooth installation experience for different users.

Documentation efforts are also observed to provide better-encapsulated installation alternatives that are easier to install, as well as versions of projects targeting different user groups. For example, in commit \href{https://github.com/SUNCAT-Center/CatLearn/commit/45e8e6cd1e5475db85174b5fc10299e16dcf69d0#diff-b335630551682c19a781afebcf4d07bf978fb1f8ac04c6bf87428ed5106870f5R50}{\textit{Example 8}}, the installation option initially provided includes obtaining the source code and performing compilation. The newly added instruction introduces an alternative approach specifically for building and running a Docker image from the Dockerfile within the source code. This provides an integrated installation experience, alleviating concerns related to compilation and other dependency management issues.

Regarding update behaviours for the installation instructions from source code, a wide range of project-specific patterns are identified. For example, modifications within code instructions covering project names, file paths for executables, and configuration file names are usually trivial, reflecting changes at the file level and repository level. A simple example in commit \href{https://github.com/akto-api-security/akto/commit/4ef6d2fb3f9b0d448c1b333df92589bc6e973635#diff-b335630551682c19a781afebcf4d07bf978fb1f8ac04c6bf87428ed5106870f5R45}{\textit{Example 9}} migrated the codebase from GitLab to GitHub, while also changing the project name from \textit{atoml-local} to \textit{catlearn}, leading to the corresponding adjustment of the commands in the README file. Updates to the source code installation option also involve adjustments of parameters, with various patterns observed for this type of change. As an illustrative example, commit \textit{Example 9} altered the Docker compose command with an additional ``\texttt{-d}'' command indicating a detached mode to run the Docker container.

\textbf{Installation-related Instructions from Published Package Operations (162)}. In contrast, users can also install software and libraries from package management systems, such as Pip, Conda, and Maven, which enable an easier installation experience. Published package installation instructions are typically not the first option OSS developers incorporate in their system, considering the additional effort required to configure the distribution binary on package management systems. In our analysed data, only 33\% of the newly added published package installation instructions do not have other installation alternatives in their previous README files. For example, in commit \href{https://github.com/mneyapo/flexx/commit/7f79092c9cf7784ed05d02a3e5a0529888d9be51#diff-b335630551682c19a781afebcf4d07bf978fb1f8ac04c6bf87428ed5106870f5R62}{\textit{Example 10}}, a new alternative was added to install the project through \textit{PyPI}, which is a package management system. However, alternative options for source code level installation were already provided in the previous README file, making the newly added published package installation option an additional choice.

Regarding the update operations in the published package installation instructions, we identified a predominant pattern that focuses on updating the project version. Trivial updates are usually found in this type of modification, with merely a version number increment for the repository's latest stable version, and sometimes combined with the development version. For example, both the stable version and the snapshot version were incremented within the code instruction for the Maven dependency of the repository in this commit \href{https://github.com/ut-parla/Parla.py/commit/ed056c29bc18159a008208f3c5456f7a742acdc8#diff-b335630551682c19a781afebcf4d07bf978fb1f8ac04c6bf87428ed5106870f5R59}{\textit{Example 11}}. 

Apart from version modification, other noticeable update operations for published package installation include changes for package names or package sources and changes for package management technology instructions. For example, the path to the image on DockerHub was updated in commit \href{https://github.com/ut-parla/Parla.py/commit/ed056c29bc18159a008208f3c5456f7a742acdc8#diff-b335630551682c19a781afebcf4d07bf978fb1f8ac04c6bf87428ed5106870f5R59}{\textit{Example 12}}, reflecting the changes in the source of the project. As an example of the change in technology instruction, in commit \href{https://github.com/osmdroid/osmdroid/commit/d4dde39a02e37c93e5552b540949d8a4a3543b41#diff-b335630551682c19a781afebcf4d07bf978fb1f8ac04c6bf87428ed5106870f5R27}{\textit{Example 13}}, the Gradle command \textit{compile} is updated to \textit{implementation} as the previous keyword was deprecated.  The rest of the updates are more specific to reflect the project development status.

\begin{tcolorbox}[left=1pt, top=1pt, right=1pt, bottom=1pt]
    \textbf{Summary of installation-related instruction results}: The concept of installation-related instruction from source code operation is the most commonly adopted choice in our analysed data. Their update behaviours vary, covering factors including project meta-data changes and parameter changes. On the other hand, the concept of installation-related instructions from published package operations is less frequent, and they are usually used as additional operations on top of the source code installation alternative.
\end{tcolorbox}

\subsection{Post-installation Instruction}

After project installation, different components are usually covered within README files. In total, 248 of 1,168 data points are related to post-installation instruction updates. We synthesised six concepts based on their tasks and discuss them as follows:

\textbf{Project Running Instruction (185)}. The most important aspect after project installation is to execute the software to accomplish specific tasks, and we observed that the vast majority of modifications in this concept fall into the running instruction changes. 

OSS projects often come with different execution options that accept different parameters as input for different use cases. In these scenarios, adding different combinations of parameters with varying but similar functionalities could lead to bloated documents, but we did notice a few instances in practice. Meanwhile, we observe that newly added commands often contain the default setting or offer a template with placeholders for parameters. For example, commit \href{https://github.com/isl-org/Open3D-ML/commit/32cd56206147f132341464df6d39a70eecb78397#diff-b335630551682c19a781afebcf4d07bf978fb1f8ac04c6bf87428ed5106870f5R135}{\textit{Example 14}} added a template execution command with two brief examples. Moreover, it included a ``get help" command, allowing documentation users to access detailed running options.

In addition, OSS projects sometimes contain multiple runnable modules, each serving a specific aspect of the project, such as server/worker modules in distributed system projects and training/inference modules in machine learning projects. Properly documenting each component's execution instruction is essential for users to make use of the software product. Within our analysed data, we identified in \href{https://github.com/uds-se/FormatFuzzer/commit/d257145ca1bd6db23f1621ed18b8e72a14057f37#diff-b335630551682c19a781afebcf4d07bf978fb1f8ac04c6bf87428ed5106870f5R45}{\textit{Example 15}} that additional execution commands were added for other blocks of functions.

\textbf{Installation Check (13)}.  After a complicated software installation process, the installation check commands allow users to verify that the software has been installed and configured successfully. In our analysed data, only a few repositories incorporated the installation check process into their documentation. Represented by commit \href{https://github.com/genixpro/kwola/commit/f5e63fcef8c60330d0c101ed87efedcbfc0b8c22#diff-b335630551682c19a781afebcf4d07bf978fb1f8ac04c6bf87428ed5106870f5R166}{\textit{Example 16}}, the added installation verification process normally contains the execution of a simple predefined code snippet. 

\textbf{Running Tests (43)}. Tests may not be required from the user's perspective, but become crucial in the context of installation in development mode. Although \textbf{Deployment Instruction (9)}, \textbf{Upgrade Instruction (6)}, \textbf{Uninstall Instruction (3)} also occurred in our analysed data, their occurrence is relatively infrequent. 

\begin{tcolorbox}[left=1pt, top=1pt, right=1pt, bottom=1pt]
    \textbf{Summary of post-installation instruction results}: The most dominant concept in this category is project running instruction. Changes in this concept are found to be related to updating parameter options and file paths to the executables. Other concepts are less frequently encountered but they all constitute the installation-related software development tasks.
\end{tcolorbox}


\subsection{Help Information}
Previous sections discuss updates of instructional commands, which focus on providing commands for the reader of the documentation to follow. In contrast, the help information serves as a resource for documentation readers to get clarification or examples to better understand the project and commands. In total, 265 out of 1,168 data points are related to the Help Information category. In this category, we synthesised three concepts, which are discussed as follows:

\textbf{Tutorial Updates (70)}. Limited in space and purpose, README files typically do not offer comprehensive usage instructions, such as API documentation. We observed that most updates within this concept involve providing code demos for the basic usage of a project. Most documentation changes are trivial, merely made to align with the source code modifications. In \href{https://github.com/MCXA/Phenotyping/commit/17d922999851c8c8f1e368be500367187ca33f2c#diff-b335630551682c19a781afebcf4d07bf978fb1f8ac04c6bf87428ed5106870f5R38}{\textit{Example 17}}, updates are made within the code demos, altering all naming paradigms for the methods of using the model.

\textbf{Explanation Information Updates (125)}. Instructional commands require users to follow step-by-step commands to complete tasks. In cases where users lack related background knowledge, they may need to blindly follow instructions. In our analysed data, we observed examples where efforts were made to enhance the accessibility of README file instructions by updating explanations.

Explanations can be added as complementary materials to different types of instruction. For example, a new dependency \textit{scikit-learn} was added in commit \href{https://github.com/SUNCAT-Center/CatLearn/commit/01bf3e9d634fdb19372756a1e7f7403aa178d667#diff-b335630551682c19a781afebcf4d07bf978fb1f8ac04c6bf87428ed5106870f5R67}{\textit{Example 18}}. However, instead of providing a simple manifest of dependencies, it includes explanations of when and why this dependency is required: \textit{If there is no interest in this level of optimisation/benchmarking, then the codebase is not needed}.  This type of information empowers documentation users to make informed decisions about whether to include a particular package when installing the project. 

Explanation is not only added where the instructions are optional, when the instructions are not common enough or potentially have side effects, additional explanation is also needed, such as the case in \href{https://github.com/SUNCAT-Center/CatLearn/commit/1e68b55808c37e2698bc082556dd7e6160bf2435#diff-b335630551682c19a781afebcf4d07bf978fb1f8ac04c6bf87428ed5106870f5R46}{\textit{Example 19}}. In this example, an explanation is added for the parameter option "\textit{no-deps}", detailing its intended outcome, along with caveats specifying when manual installation of dependencies for other modules may be necessary.

\textbf{Notes and Troubleshooting (70)}. In our analysed data, we identified documentation efforts in updating notes and troubleshooting content. A brief definition of ``notes'' is \textbf{additional} information or important considerations that must be taken into account when following the instruction. Meanwhile, troubleshooting is dedicated to providing guidance towards common errors that people may encounter. 

For example, the commit \href{https://github.com/awslabs/aws-crt-java/commit/dad3837984c3bfe4e260dedd4196a95dc9b88d17#diff-b335630551682c19a781afebcf4d07bf978fb1f8ac04c6bf87428ed5106870f5R99}{\textit{Example 20}} added a note, specifying additional steps that need to be done for \textit{linux-armv6} and \textit{linux-armv7} systems. On the other hand, in commit \href{https://github.com/decentralized-identity/universal-resolver/commit/81553bd600bf31cf61a3954c16fe54a4e3f33e20#diff-b335630551682c19a781afebcf4d07bf978fb1f8ac04c6bf87428ed5106870f5R51}{\textit{Example 21}}, troubleshooting information was added, including a description of the problem related to an outdated docker-compose version and the solution provided to address this issue. These instructions are crucial for specific user groups, and the inclusion of common notes and troubleshooting information can enhance users' experiences by reducing the time spent searching for solutions.

\begin{tcolorbox}[left=1pt, top=1pt, right=1pt, bottom=1pt]
    \textbf{Summary of help information results}: The most frequent updates in this category are the explanation information updates, followed by notes and troubleshooting updates and tutorial updates. Unlike instructional updates, updates in this category focus more on providing clarifications and examples, but are also less frequent.
\end{tcolorbox}


\subsection{Document Presentation}
Different from the previous sections, document presentation updates involve less content modification. Instead, the emphasis is on organising the documentation to enhance visual presentation and reading comprehension. In total, 477 of 1,168 data points are associated with this category, indicating a relatively heavy documentation effort. In this category, we synthesised five concepts, which are discussed as follows:

\textbf{Documentation Structural Modification (197)}. The structure of the documentation contains essential information for the reader to distinguish its source, topic, and other characteristics~\cite{bratko2006exploiting}.  In our analysed data, we identified a large amount of effort in document structure updates, reaching around 16\% of all data.

\begin{figure}[ht!]
    \centering
    \begin{subfigure}[b]{\columnwidth}
         \centering
         \includegraphics[width=\columnwidth]{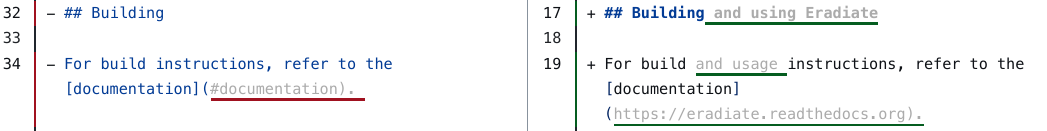}
         \caption{}
         \label{fig:Section-name-update-1}
     \end{subfigure}

    \begin{subfigure}[b]{\columnwidth}
         \centering
         \includegraphics[width=\columnwidth]{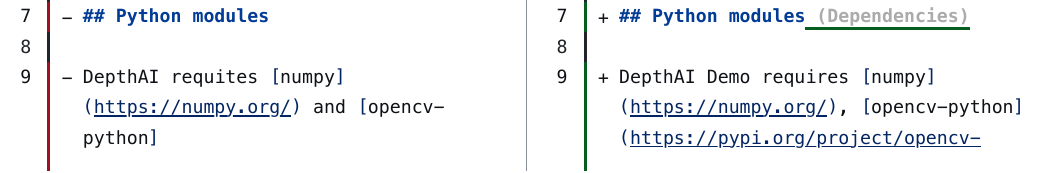}
         \caption{}
         \label{fig:Section-name-update-2}
     \end{subfigure}

     \caption{Section Header Name Update Examples}
     \label{fig:Section-name-update}
\end{figure}

Section headers are indicators of the content within the sections. Figure~\ref{fig:Section-name-update} shows examples for updating the section names to more accurately reflect the content under the section. Figure~\ref{fig:Section-name-update-1} updates the header name as a result of the inclusion of new distinct contents in the section, while Figure~\ref{fig:Section-name-update-2} undergoes an update with the aim of improving the clarity of the header of the section. 

In addition, when a section is cluttered with excessive contents, it becomes necessary to re-organise the materials in a more logical manner by relocating them into other sections. In commit \href{https://github.com/itsallcode/white-rabbit/commit/bd1cefc49d728b8bbcbdabc0df169d9a44491784#diff-b335630551682c19a781afebcf4d07bf978fb1f8ac04c6bf87428ed5106870f5}{\textit{Example 22}}, the instructions in the previous README file are largely gathered in a single section. The updates, as indicated in the commit, distribute the corresponding parts among different subsections, enhancing the organisation of the documentation for better comprehension. This also happens when instructions are introduced with the same purpose but with different technologies. For example, in commit \href{https://github.com/actinia-org/actinia-core/commit/4f7363b74033a40b3265e1cfc970bb2ee59bea55#diff-b335630551682c19a781afebcf4d07bf978fb1f8ac04c6bf87428ed5106870f5R47}{\textit{Example 23}}, the Docker installation instruction is moved under ``Installation with Docker'' after adding the Pip install instruction. 

Other modification operations including adjusting the section header size, section hierarchical changes which organise multiple sections under a single common section, and section order updates are less commonly found in practice, but all contribute to the organisation of information in README files.


\textbf{Content Formatting and Cleaning (119)}. README content formatting involves using Markdown syntax to either highlight or organise the content. Various information is commonly formatted, including large code blocks in \href{https://github.com/jupyterhub/dockerspawner/commit/4a90a55e8449d3d303e96610e41b2ef137b51811#diff-b335630551682c19a781afebcf4d07bf978fb1f8ac04c6bf87428ed5106870f5R69}{\textit{Example 24}}, inline code blocks in \href{https://github.com/Tolu-gith/tt/commit/907705873708227d568a72d5e3b200713c6b14d4#diff-b335630551682c19a781afebcf4d07bf978fb1f8ac04c6bf87428ed5106870f5R24}{\textit{Example 25}}, and URLs in \href{https://github.com/alorence/django-modern-rpc/commit/d2d028aae74e328c59aae06e8142c2b67cb1d54a#diff-b335630551682c19a781afebcf4d07bf978fb1f8ac04c6bf87428ed5106870f5R36}{\textit{Example 26}}. All of these formatting operations contribute to enhancing the presentation of the documentation, making the README files easier to follow. Specifically, the formatting on code blocks enables better highlighting of code components, while the formatting on URLs facilitates enhanced accessibility and clearer content display. Plain text is also formatted to highlight important information, as indicated in \href{https://github.com/darkwizard242/cis_ubuntu_2004/commit/c26a1c99a1427d1cfe0985b7443a27f0d99ce918#diff-b335630551682c19a781afebcf4d07bf978fb1f8ac04c6bf87428ed5106870f5R379} {\textit{Example 27}}, with bold and italic texts.

Other formatting includes adding empty lines to separate logically different paragraphs, using bullet points, or an ordered list to organise the instructions within the documentation. These formatting operations organise the content within a section in a more logical manner to improve the reading experience. 

\textbf{Presentation Fixing (61)}. Presentation fixing in the domain of README documentation presentation involves addressing two perspectives: grammar and typo fixing, as well as Markdown syntax fixing.

Errors in grammar and typography can influence the reading experience, but may not severely impact the documentation's usability. In commit \href{https://github.com/PMCC-BioinformaticsCore/janis-core/commit/c72096c153e94f4c65709e02ee7c36606cccf61b#diff-b335630551682c19a781afebcf4d07bf978fb1f8ac04c6bf87428ed5106870f5R6}{\textit{Example 28}}, the whole commit is dedicated to fixing typos to improve readability. For example, the word \textit{Pyhton} is corrected to \textit{Python}. On the other hand, Markdown syntax errors sometimes prevent the content from correctly rendering, making the documentation difficult to read. For example, in commit \href{https://github.com/MaximProkhorov/vera-pdf-library/commit/1a86148b0dc154eb14c2dfd14a07fe92a1ee747f#diff-0324feccee544c74b7e269abb8ef3840e485a44db47736ad2ac3d8d658b415edR9}{\textit{Example 29}}, an extra space is inserted between the hash symbol and section header, ensuring the correct display of the header.

\textbf{Text Rephrasing and Editing (70)}. Text rephrasing and editing is different from updates in instructions or help information, as the latter two entail different or additional information in the actual guidance, procedures, or technical details, while the first one focuses on modification of the wordings and sentence structures. 

Regarding the style of text rephrasing and editing, we observed instances where efforts were made to simplify the sentences or render them more succinct. In commit \href{https://github.com/actris-cloudnet/cloudnetpy/commit/458623df6af9bd84451f291b2dff256120d2c265#diff-b335630551682c19a781afebcf4d07bf978fb1f8ac04c6bf87428ed5106870f5R50}{\textit{Example 30}}, the phrase \textit{instead of `pip install .`} was deleted. Since there already exists the correct code command, providing an incorrect command is unnecessary and could potentially lead to users' misuse. On the other hand, there are also light updates that perform less apparent modifications. For example, in commit \href{https://github.com/Ravaelles/Atlantis/commit/89fe47619edf026500c1db491679bd4fd3b8d7f7#diff-b335630551682c19a781afebcf4d07bf978fb1f8ac04c6bf87428ed5106870f5R25}{\textit{Example 31}}, the phrase \textit{doesn't necessarily work} is revised to \textit{might not work as expected}. The modified text included the behaviour that the code does not work, highlighting the potential limitations.

\textbf{Self-admitted Documentation Debt Management (30)}. Although uncommon, there are projects on which developers explicitly leave unfinished marks in certain sections. In commit \href{https://github.com/Kuifje02/vrpy/commit/bd6b882db0fe99f6a78570d94312fe6277293061#diff-b335630551682c19a781afebcf4d07bf978fb1f8ac04c6bf87428ed5106870f5R43}{\textit{Example 32}}, an installation section is added, with only a placeholder ``coming soon". Although in most scenarios, these self-admitted documentation debts are removed once the contents are added, we identified cases in which the documentation debt is only commented out using Markdown syntax. In this case, the reader can still access the complete information, while the documentation maintainers are aware of the outstanding documentation debt from README files. 

\begin{tcolorbox}[left=1pt, top=1pt, right=1pt, bottom=1pt]
    \textbf{Summary of document presentation results}: Documentation structural modification includes changes in section headers, and organising content underneath them. The concept of content formatting and cleaning involves using Markdown syntax to either highlight or organise the content. Error fixing includes both Markdown syntax fixing and grammar and typography fixing. In addition, the concept of text rephrasing and editing is primarily related to making text simpler and more succinct. Furthermore, we also identified a few cases where self-admitted documentation debt is managed within README files.
\end{tcolorbox}

\subsection{External Resources Management}
External Resources refer to other documentation and non-documentation artefacts that are not embedded within the project's README files but are accessible only via the provided URL links. In total, 366 of 1,168 data points are related to this category. In this category, we synthesised three concepts, which are discussed as follows:

\textbf{Project External Documentation Management (120)}. README files serve as a starting point for an OSS project, which contains essential information for users to learn and start on a GitHub repository~\cite{wang2023study}. However, including all information within the README files would lead to information overload, hindering documentation users from quickly identifying their desired contents. In our analysed data, we discovered one common activity as managing the project's external documentation, taking up around 10\% of the whole data instances.

In commit \href{https://github.com/agonyforge/arbitrader/commit/ee31e68e547e747dafbfd443e7718ebc373c73b8#diff-b335630551682c19a781afebcf4d07bf978fb1f8ac04c6bf87428ed5106870f5R12}{\textit{Example 33}}, the content regarding ``how-to-use'' became overly lengthy and complicated. This commit reorganised the content into a Wiki document for this project, with a URL provided to easily navigate to this external documentation. Meanwhile, it also provided a URL redirect to another Markdown file with contribution information. Generally, we found that the newly introduced documentation has a wider range of formats, including Markdown files, HTML files, Wiki files, RST files, and so on. There are also update operations for the URL of external documents, which can be rooted to change of documentation location or new version reflection. In fact, we observed that the addition of external documentation resources is often accompanied with the deletion of instructional content within the README, which corresponds to documentation organisation efforts.

\textbf{Project Non-documentation Artefacts Management (136)}. Apart from external documents, other artefacts (e.g., source code, release package, data, and configuration file) are closely related to installation-related tasks. Typically facilitated by Git, the most frequent modification for the non-documentation artefact is on source code access. This access is a crucial step to enable the source code installation method. Release packages, such as the jar package for Java projects, are also actively maintained for users to get access to different versions of the release. Additionally, some projects organise other URLs, such as links to configuration files (e.g., Dockerfile), providing quick access to corresponding files, such as the update in commit \href{https://github.com/sdauzcm/SR-basicSR/commit/de3ba707e538aa1d8395376a9fae5e20c9c4ef40#diff-b335630551682c19a781afebcf4d07bf978fb1f8ac04c6bf87428ed5106870f5R44}{\textit{Example 34}}. This practice provides quick access to editable files, which helps with local development.

\textbf{Third-Party External Resources Management (110)}. Although third-party requirements are important for OSS projects, it is not the responsibility of the README file to provide detailed information for them; doing so may lead to issues of tangled information~\cite{uddin2015api}. Therefore, we identified that the majority of data instances in this concept involve adding or updating URLs for third-party dependencies or other requirements of this project, such as the update in commit \href{https://github.com/BlueBrain/NeuroMorphoVis/commit/f32210d72a4ac9141b3f5b5af6e8231d0de7ef4c#diff-b335630551682c19a781afebcf4d07bf978fb1f8ac04c6bf87428ed5106870f5R13}{\textit{Example 35}}. These changes often accompany related instructional changes, and the URLs typically lead to their official websites or documentation sources. 

\begin{tcolorbox}[left=1pt, top=1pt, right=1pt, bottom=1pt]
\textbf{Summary of external resources management results}: Project external documentation management is related to providing links for other detailed project documents, which helps avoid lengthy README files. Project non-documentation artefacts refer to URLs for project artefacts including source code, executables and other configuration files. Furthermore, the concept of third-party external resources management corresponds to updates for URLs for external requirements of the software project.
    
\end{tcolorbox}

\subsection{Cross-Cutting Concerns}
There are several perspectives that we identified during the analysis that are scattered across different categories and important to the README file update operations. Therefore, we dedicate a subsection to discuss them.  

(1) \textbf{Version Management}. 
As described in previous sections, version management is a notable area where substantial efforts are invested when updating README files. With the development of the project, new versions are released to incorporate additional features, which leads to the active update for version information. In summary, we identified that version management primarily focuses on project versions, dependency versions, and operating system versions.

In terms of the project version, information is provided for the latest stable version and the developing version. This version information is usually updated within installation-related instructions, project running instructions, URLs for accessing project artefacts, and plain text manifest. Additionally, previous versions are sometimes provided, and outdated versions are included as help information. Regarding dependency version updates, the programming language version will be updated when a newer version is incorporated into the software project, e.g., migrating from Python 2 to Python 3 or adding support for the latest version of JVM. As for other dependencies, although version information is not always displayed in the README file for every repository, their versions are also updated accordingly in the case that they are displayed, to reflect the changes within each project. As indicated in the previous example, operating system version updates reflect project changes that introduce a newer compatible version. In addition, when a new compatible operating system is added to the project, it might be necessary to document specific instructions on different software development tasks.


(2) \textbf{Development related Instruction Updates}. Development-related instructions require different settings from normal usage mode, and are documented in README files to facilitate better contribution guidelines.

Starting from pre-installation setup, the development mode requires slightly different or additional dependencies compared to the normal usage mode. For example, \textit{pytest} is only useful for testing and is not required for normal use cases. In terms of installation-related processes, separate instructions are usually added for development mode, with an additional flag in the command parameter to handle extra procedures that are needed. Moreover, some projects also provide a snapshot version of the project that is under development for users to experience. Regarding post-installation instructions, development-specific activities are included to complete the development guidelines, which includes testing and processes to contribute to the project. 


(3) \textbf{Error Fixing}. Errors are in various forms, as summarised above. The impact on documentation users varies depending on the nature of these errors. Document presentation errors are among the least severe, mainly affecting the readability of README files. In addition, presentation errors are relatively easier to solve by applying additional grammar and Markdown syntax checking tools.

However, we also observed efforts to address instructional errors. Instructional errors are scattered across pre-installation, installation-related, and post-installation sections, and their presence can lead to failure of command execution, thus hindering the usability of README files. Errors also exist in URLs that serve as links to external resources. These types of errors are less apparent and more severe, requiring detailed examination or even execution to identify and a higher level of expertise to rectify.

\subsection{README Update Behaviours}

Last, we summarise the overall update behaviours based on the derived codes. The most frequently updated category is "Pre-installation instruction," with dependency setup being the most modified concept. Documentation presentation is the second most focused category, where maintainers primarily work on modifying the documentation structure and formatting the documents for better readability. Additionally, updates to instructional commands significantly outnumber updates to help information, with nearly five times as many. These observations indicate that OSS documentation maintainers invest more effort in updating dependencies and documentation structures. They also focus more on providing instructional information rather than help information such as troubleshooting in README files.

Additionally, we distinguish our base codes with the operations of ``Addition'', ``Modification'', and ``Removal'', with the statistics displayed in Table~\ref{tab:taxonomy}. Unlike the section-level updates shown in Figure~\ref{fig:header-update-bar}, we categorise them based on more detailed modification behaviours. Specifically, whether the modification introduces new information, updates existing information, or removes old information. Seen from the table, apart from document presentation, the largest proportion of update behaviour involves ``addition''. This indicates that more information is being incorporated into README files as the software development process progresses, requiring documentation maintainers to spend time assembling new information to reflect system changes.

\begin{table*}[t]
        \centering
        \caption{Quantitative evidence on update categories and information presented}
        \resizebox{0.9\textwidth}{!}{
        \begin{tabular}{lrrrrr}
        \toprule
                          \multirow{2}{*}{\textbf{Update}}  & \multicolumn{5}{c}{\textbf{Information}} \\
        \cmidrule(l){2-6}
                                                & Pre-installation & Installation & Post-installation & External resources & Help information \\
        \midrule                                        
        Pre-installation instruction     & 27                       & 19                   & 14                        & 17                         & 11                       \\
        Installation-related instruction  & 15                       & 28                   & 17                        & 18                         & 7                        \\
        Post-installation instruction     & 19                       & 26                   & 29                        & 10                         & 10                       \\
        External resources management    & 20                       & 26                   & 15                        & 26                         & 9      \\     
        Help information                 & 16                       & 24                   & 20                        & 15                         & 20                       \\
        \midrule
         Document presentation            & 16                       & 30                   & 17                        & 13                         & 7                        \\
        \bottomrule
        \end{tabular}
        }
        \label{tab:sec-cat-co}
    \end{table*}

We also provide quantitative evidence on information already present in the first version of a README when different categories of updates occur. Specifically, we sampled 30 documentation updates from each category and annotated the first version of each README to identify the occurring information. As approximately 60\% of the data involves only one type of update category, and to better illustrate the distribution of sections across independent update behaviours, we limited our sampling to data containing only a single update category. Table~\ref{tab:sec-cat-co} shows the frequency of each section present in the previous README version (totalling 30 samples) when each follow-up modification occurs, highlighting the relationship between section completeness and modification patterns.

From the table, it is evident that all types of information generally exist in the initial version of a README file. Notably, in the first version across various update behaviour categories, installation information is the most frequently included content, except in cases where pre-installation instructions are updated. Moreover, for each update behaviour category (except for document presentation), related content was already present in the initial version of the README. This finding suggests that practitioners often update documentation incrementally rather than introducing entirely new content.

\section{Triggers for Documentation Updates}

Additionally, we analysed the factors that triggered documentation updates across the entire set of 1,163 documentation updates. Specifically, we followed a series of steps to identify these triggers, stopping once the trigger was identified or reaching the final step. The steps are as follows: (1) read and understand the document update, (2) review the previous README version for obvious issues (e.g., syntax errors, incomplete instructions, presentation issues), (3) examine the commit message to determine whether it is directly linked to a pull request or issue, or if it includes the motivation for the update, (4) if the commit is associated with a pull request/issue, review the first block of the description and initial discussion of that pull request, (5) examine any additional pull requests linked in one level in the discussion thread (non-recursive), (6) check the most recent release for the repository corresponding to this README version, (7) search for modified keywords such as package names and file paths using the GitHub search function, and (8) if none of the above methods yields an answer, we mark the trigger as ``no explicit evidence'' as we were not able to identify explicit evidence for what triggered the update.

This procedure for identifying triggers was developed based on an initial set of pilot data, discussed and refined until reaching consensus among all authors. The design of this process aims to identify triggers using the most commonly used channels in open-source projects while avoiding spontaneous and endless searching, ensuring that our method is replicable. After that, the first author analysed the rest of the triggers and flagged uncertain entries for further discussion.

Following this procedure, we identified 614 triggers from the entire 1,168 documentation updates, while the rest of the documentation updates are annotated with ``no explicit evidence''. We then performed open card sorting~\cite{zimmermann2016card} to group similar triggers into the same group and assign meaningful names to it. Specifically, we identified 11 different types of triggers. These 11 triggers are further grouped into 3 bigger themes, namely 1) Errors in the previous documentation, 2) Changes in the codebase, and 3) Need for documentation improvement. In the following, we introduce each type of trigger, indicating in parentheses how many instances we identified for each type.

\subsection{Errors in the previous documentation}

\textbf{Presentation issues in the previous version of README (93).} Documentation maintainers update their documentation when identifying issues within the previous version of a README that impact its presentation. For example, the commit~\cite{githubMissingWhitespaces} added a white space between the hashtag and the section title, as the previous version of the README violated the markdown syntax for displaying headers.

\textbf{Incomplete contents in the previous version of README (78).} In this case, documentation maintainers left the documentation clearly unfinished in the previous version. Documentation debt, such as explicit \textit{TODOs}~\cite{githubTODO} or documentation with only skeletons of section headers~\cite{githubSkeleton}, are strong indicators of unfinished jobs, suggesting that new content should be added in accordance with the context provided. Meanwhile, when the README document is added to the project at the beginning (i.e., the first version of the README), a lot of content is needed, and our proposed template might be a good starting point.

\textbf{Instruction or content issues in the previous version of README (48).} Instruction or content errors severely impact downstream users' experience in installing and using the project, and developers should quickly identify the existence of such issues and fix them. For example, commit~\cite{githubCommand} removes command \texttt{--system-site-packages}. The error in this command is reported in issue 1 in this repository by a user and its fix is later reflected in the document.

\textbf{Outdated elements in previous README (12).} Software documentation gets silently outdated during the software development process. When developers realise the existence of such outdated pieces, they make adjustments in the README documents accordingly. For example, commit~\cite{githubOutdated} updates the Python version to 3, while the migration from Python 2 to Python 3 was done in a pull request three years before this documentation update~\cite{githubMigration}, indicating the existence of such an outdated element in the README file.

\subsection{Changes in the codebase}
\textbf{Update of software version (131).} This is the most frequent trigger. When new releases of a software are made, which can usually be observed in the repository release channel or release log, version information in the documentation gets incremented as well, as showcased in this example~\cite{githubRelease}.

\textbf{Add or modify software functionality (83).} This relates to large changes in the codebase, usually accompanied by adding, updating, or deleting multiple files. As a result of such updates, software functionalities such as installation or running instructions might be updated. This requires the documentation authors to carefully account for the changes to be updated in the README. For example, pull request~\cite{githubPRFunc} made large updates with various functionalities, and the new OS support is reflected in the README within the commit~\cite{githubRMFunc}. Such information is difficult to summarise from the large amount of code changes, while clear pull request descriptions can help developers easily track what should be documented in the README files.

\textbf{Command updates or file path updates (76)}. Simple command and file path updates are short code snippet modifications that can be easily tracked to the corresponding files and directories. When such code elements get updated during the development, README changes are also required to keep the documentation up-to-date. For example, commit~\cite{githubSnippetChange} moves the directory \texttt{configs} from the root folder to the \texttt{code2seq} folder, thus triggering updates to the file path present in the documentation.

\textbf{Update of software dependencies (52).} Dependencies evolve as the software evolves. Updates to dependencies also triggers documentation updates. For example, commit~\cite{githubDependencyRM} updates the dependency Confluent version to 3.1.2, while such an update can be identified within another commit~\cite{githubDependencyC} within the same pull request to update the \texttt{build.gradle} file. In fact, the dependency update information can usually be found within the build file or requirement file in the same repository.

\subsection{Need for documentation improvement}

\textbf{Self-motivated documentation improvement (50).} Some projects inspect their README document periodically to evaluate its quality and propose updates to make it easier to understand. For example, a lot of effort was spent in pull request~\cite{githubSelfMotivate} to make the README file better. Incorporating our proposed template could help mitigate the effort spent on this aspect.

\textbf{Request of documentation updates by downstream users (13). } Sometimes downstream users proactively approach the developers, requesting an explanation or enhancement of the documentation. The GitHub Issue channel is suitable for monitoring such concerns raised by users. For example, issue~\cite{githubIssue} asked for an installation option on Ubuntu and Debian, and the instruction was added later in commit~\cite{githubIssueAdd}.

\section{README Template and its Validation}

\begin{figure}[t!]
    \centering
    \includegraphics[width=\columnwidth]{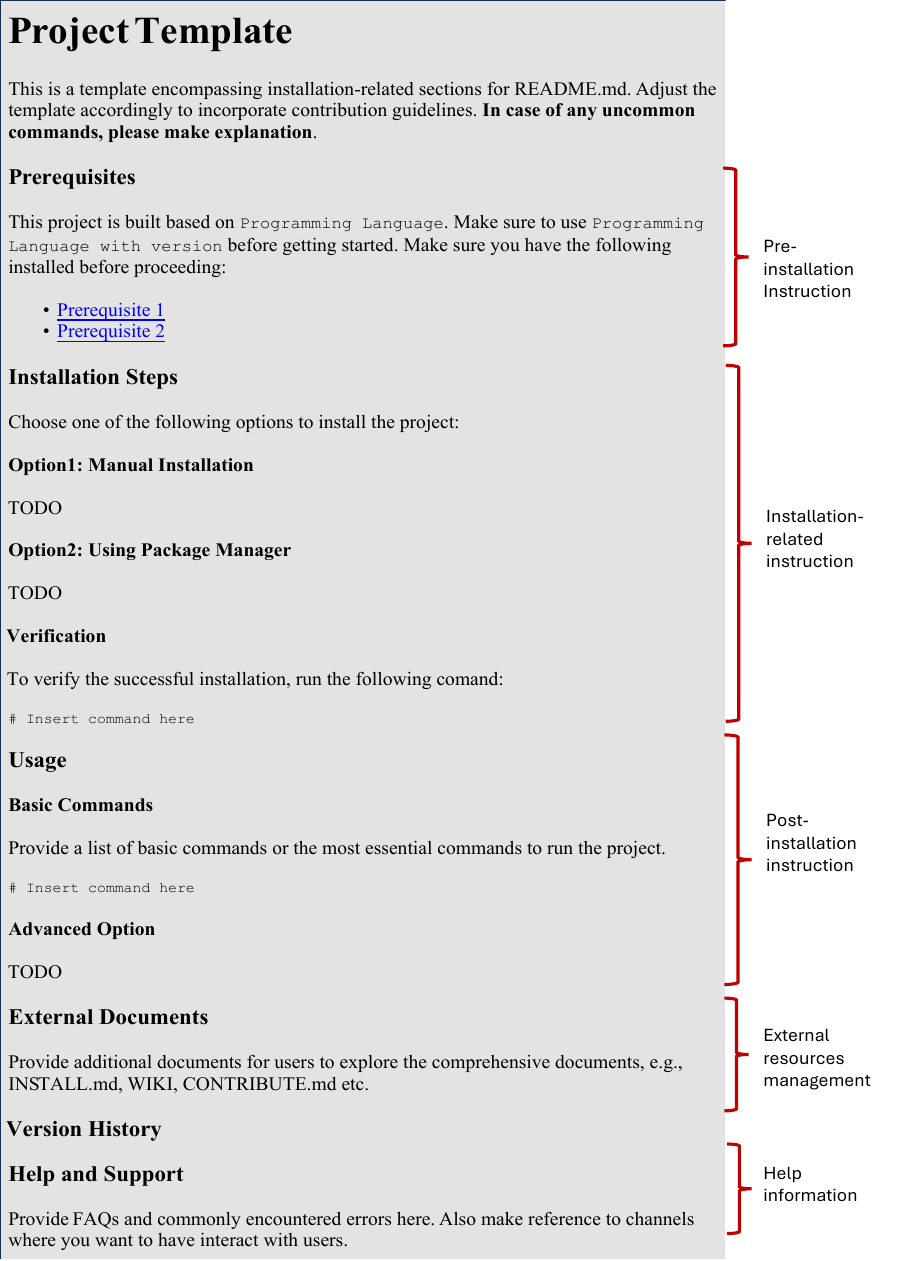}
    \caption{Proposed README Template}
    \label{fig:template-md}
\end{figure}

As we discovered that most categories contain more addition operations and that significant efforts are spent on documentation presentation adjustments by practitioners, we created a README template based on the derived taxonomy. This template encompasses all essential installation-related contents, aiming to provide a comprehensive checklist and reduce practitioners' maintenance efforts for README files. In this section, we describe our proposed template and validate it through a survey conducted on Prolific.

\subsection{README Template Content}


Figure~\ref{fig:template-md} presented the suggested template. It covers basic instructions including prerequisites, installation steps, usage, external documents, version history, and help and support sections. These sections are linked to the taxonomy derived from our RQ1, and we have explicitly labelled them in the figure. In the prerequisites section, we incorporate programming language version and dependencies for the project. URLs for the official websites of the dependencies are also provided. The installation steps include both source code level installation and published package level installation. The usage section covers basic commands for running the project, as well as some advanced options to display the capability of the software. External documents provide links to external comprehensive documents. The version history section provides the version information of this project, while the help and support section provides most common FAQs and channels for users to interact with.  In cases where specific functionalities are yet to be implemented, these placeholders can be seen as a form of self-admitted document debt, serving as reminders for document maintainers to complete the necessary information.

\subsection{Template Validation Process}

To further verify whether our proposed template is perceived as useful by the documentation users, we conducted a survey on Prolific. We applied our proposed template to the original README documents by performing rewrites. To avoid subjectivity in the rewriting process, we instructed ChatGPT to handle this task. Specifically, we provided the original README file and the template, along with two additional instructions: (1) do not omit any details from the original document, and (2) do not add any information not present in the original document. We used ChatGPT 4o throughout this procedure. Only one entry of instruction is provided to ChatGPT to collect the generation, without further tuning or additional feedback.

For the evaluated samples, we first randomly selected 20 repositories. From each repository, we randomly chose one commit to ensure each document was selected only once per repository. We then instructed ChatGPT to apply our proposed template to these documents. We refer to the original documents as ``original" documents and the generated documents as "augmented" documents. In total, we obtained 20 ``original" to ``augmented" document pairs for evaluation.  

Table~\ref{tab:header-count-sample} shows the number of projects containing the top two levels of headers in both the ``original'' and ``augmented'' documentation. For these two most frequent headers, the majority of original documents feature these top two header levels, indicating that the original README versions are already structured. The augmented version, however, contains slightly fewer top-level headers. As a result, the evaluation is not biased by having more structure in the augmented version, allowing for a focus on the content and organisation of our proposed template.

 \begin{table}[t]
    \centering
    \caption{Counts for Header Scales for Sampled Repositories}
    \begin{tabular}{c c c}
    \toprule
        Header Scale &  $\langle h1 \rangle$ & $\langle h2 \rangle$  \\
        \midrule
        Original Version Count  & 18 & 16  \\
        Augmented Version Count & 13 & 15 \\
    \bottomrule
    \end{tabular}
    \label{tab:header-count-sample}
\end{table}

We conducted a Prolific survey by dividing the 20 document pairs into 10 different surveys, with each survey taken by three different participants. We performed a sequential survey release as in~\cite{gao2023evaluating} to prevent the same individual from participating multiple times. Meanwhile, we followed a study design similar to Nadi and Treude~\cite{nadi2020essential} by inserting a ``quality gate'' in the survey. After reading the documents, we asked the participants to answer a simple question, e.g., ``Is Gradle used in the above-mentioned project?''. We filtered out participants who did not answer the question correctly.

To evaluate how users perceive the documents, we adopted an evaluation framework proposed by Treude et al.~\cite{treude2020beyond}, which provides a ten-dimensional framework for assessing software documentation quality.  Table~\ref{tab:quality-dimension} displays these dimensions as well as questions in this paper. At the beginning of the survey, we first provided the definitions of these ten dimensions and asked the participants to read through them. After that, they were requested to read through the original and augmented documents and provide scores on these ten dimensions. To prevent biased judgements, the documents were anonymised regarding their status as original or augmented. Likert scale is used to mark each of the aspects, with a score of 1 for strongly disagreeing and a score of 5 for strongly agreeing. Additionally, we revealed the template at the end of the survey and with an open-ended question aiming to gather participants' opinions about it. We added a sample as well as all README file pairs in our replication package.

\begin{table}[]
    \centering
    
     \caption{Dimensions of software documentation quality}
     \resizebox{\columnwidth}{!}{%
     \begin{tabular}{l l}
     \toprule
         Dimension&  Question\\
         \midrule
         Quality& How well-written is this document (e.g., spelling, grammar)\\
     Appeal&How interesting is it?\\
     Readability&How easy was it to read?\\
     Understandability&How easy was it to understand?\\
     Structure&How well-structured is the document?\\
     Cohesion&How well does the text fit together?\\
     Conciseness&How succinct is the information provided?\\
     Effectiveness&Does the document make effective use of technical vocabulary?\\
     Clarity&Does the document contain ambiguity?\\
     \bottomrule
     \end{tabular}
     }
   
    \label{tab:quality-dimension}
\end{table}

\begin{figure*}[t!]
    \centering
    \includegraphics[width=\textwidth]{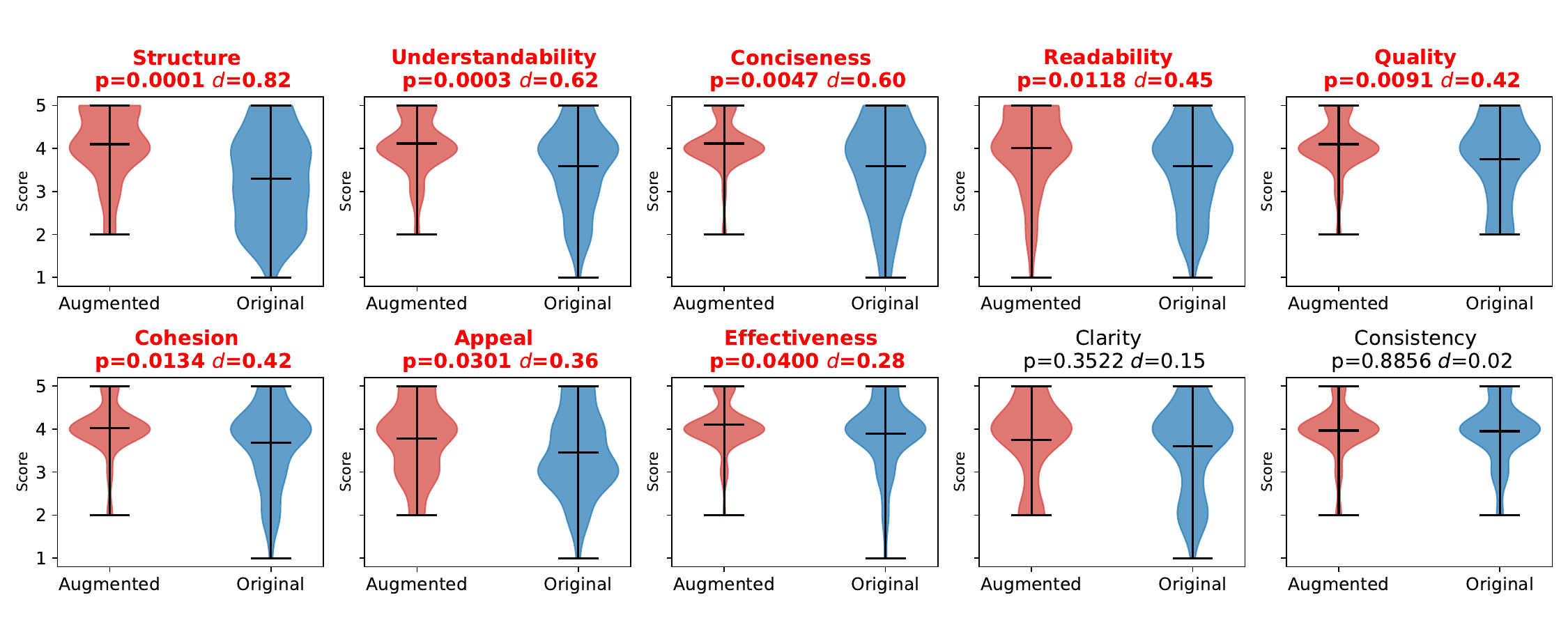}
    \caption{Survey Result}
    \label{fig:survey-result}
\end{figure*}

\subsection{Demographics of Annotators}
We first applied predefined filters on Prolific, requiring participants to be in the Information Technology (IT) employment sector and be proficient in English. We asked the same question about their employment again at the beginning of our survey and filtered out 18 participants who did not answer this question consistently with their Prolific registration. Our ``quality gate'' question further filtered out four participants.

\begin{figure}[t]
    \centering
    \includegraphics[width=\linewidth]{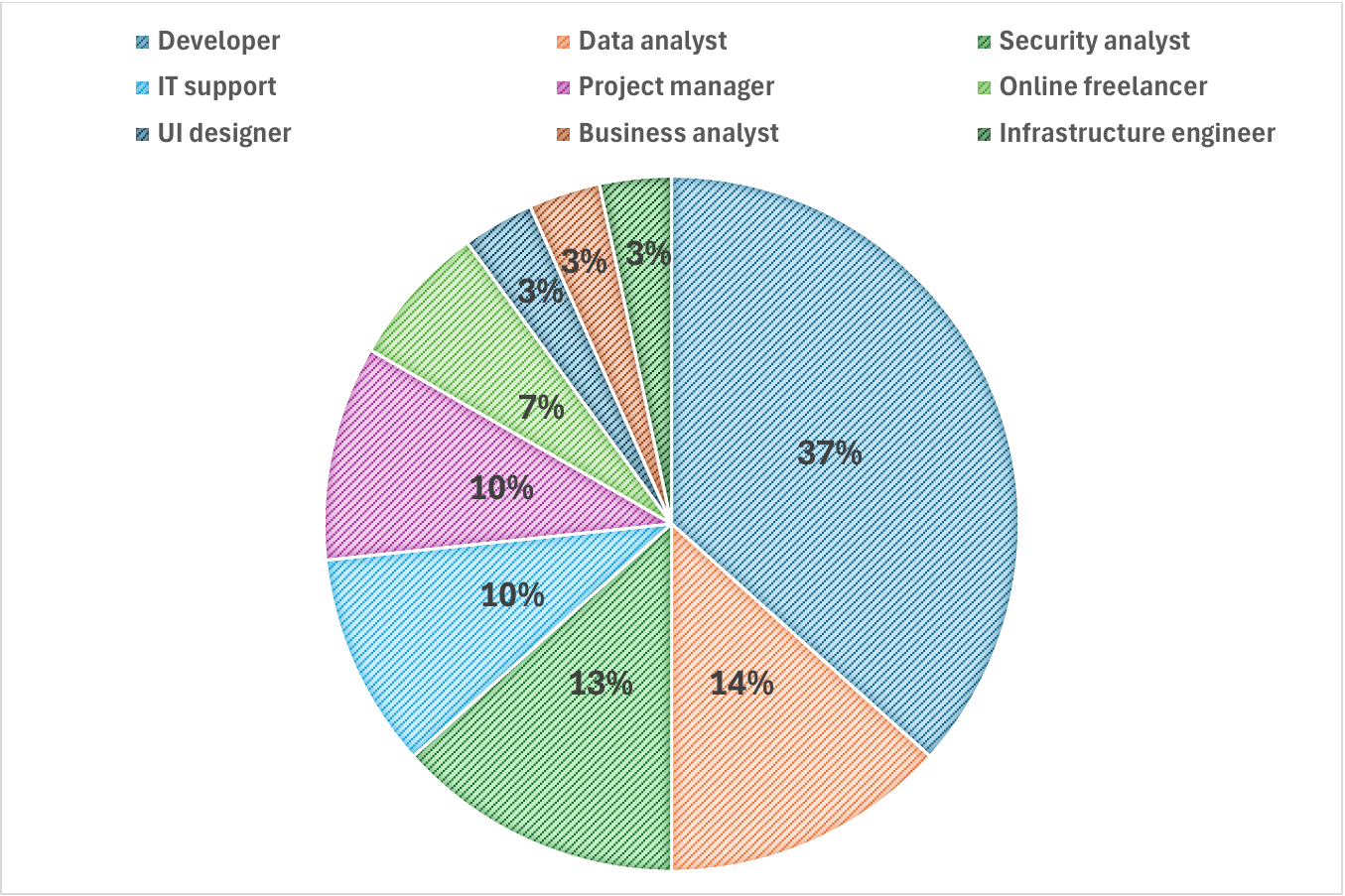}
    \caption{Job Roles of Annotators}
    \label{fig:demographics}
\end{figure}

In the survey, we asked participants about their job roles and how many years they have worked in the IT field. For the 30 participants who passed all filters and submitted their responses, they have on average 6.9 years of experience working in IT, with a maximum of twenty years and a minimum of one year. Figure~\ref{fig:demographics} shows the demographic distribution, with the largest proportion of participants being ``developers'', and the legend sorted in descending order based on the number of participants. We computed the inter-rater agreement among the evaluators, which reached 0.34 for Krippendorff's alpha~\cite{krippendorff2004reliability} coefficient overall. This indicates that the evaluators reached moderate agreement, while the evaluation remains somewhat subjective.

\subsection{Template Validation Result}

Seen from Figure~\ref{fig:survey-result}, the average scores for the ``augmented'' document are consistently higher than those of the ``original'' documents. The figure is sorted based on the difference between the average scores of these two types of documents. The most improved dimensions are structure, understandability and conciseness, with a gain of 0.80, 0.53, and 0.53 of the Likert average score, respectively. Meanwhile, the increases in clarity and consistency are less noticeable.

We also performed Wilcoxon signed-rank tests~\cite{woolson2005wilcoxon} on all dimensions to determine whether the differences are statistically significant. As indicated in the figure, eight out of ten dimensions are with a p-value below 0.05, indicating statistically significant differences. We further computed the Cohen's d effect size~\cite{becker2000effect} for the ten evaluation dimensions. The results indicate that the adoption of our proposed template significantly enhances the perceived structure with a large effect, while significantly enhances understandability, conciseness, readability, quality, cohesion, appeal, and effectiveness with medium effect. The improvements in clarity and consistency are small.

Therefore, based on the quantitative results of our survey, we find that applying our proposed template increases the quality of the generated documents, particularly in terms of structure, understandability, and conciseness, without compromising clarity and consistency, as perceived by document readers.

\subsection{Practitioners' Perspectives from Pull Requests}

To further assess the proposed README template and methodology, we conducted an additional study to gather developers’ opinions. Specifically, we applied the proposed template to the latest version of README files and submitted pull requests to the respective repositories, requesting the suggested changes. Out of the 400 repositories initially considered, we excluded those without any commits in the past year, narrowing the sample to 219 active repositories. From these, we randomly selected 30 repositories to conduct the experiment.

In the pull request, we included a commit with the updated README file that integrates our proposed template. Along with this, we provided a description of the potential benefits of adopting the template and included a screenshot of the raw template for reference. At the time of writing, among eight repositories that responded and were comfortable with LLM-generated improvements to their README files, six projects responded positively, while two projects responded negatively. Four repositories expressed reluctance to use LLM-generated content for documentation or questioned the purpose of unsolicited pull requests.

\begin{figure}[t!]
        \centering
        \includegraphics[width=0.9\linewidth]{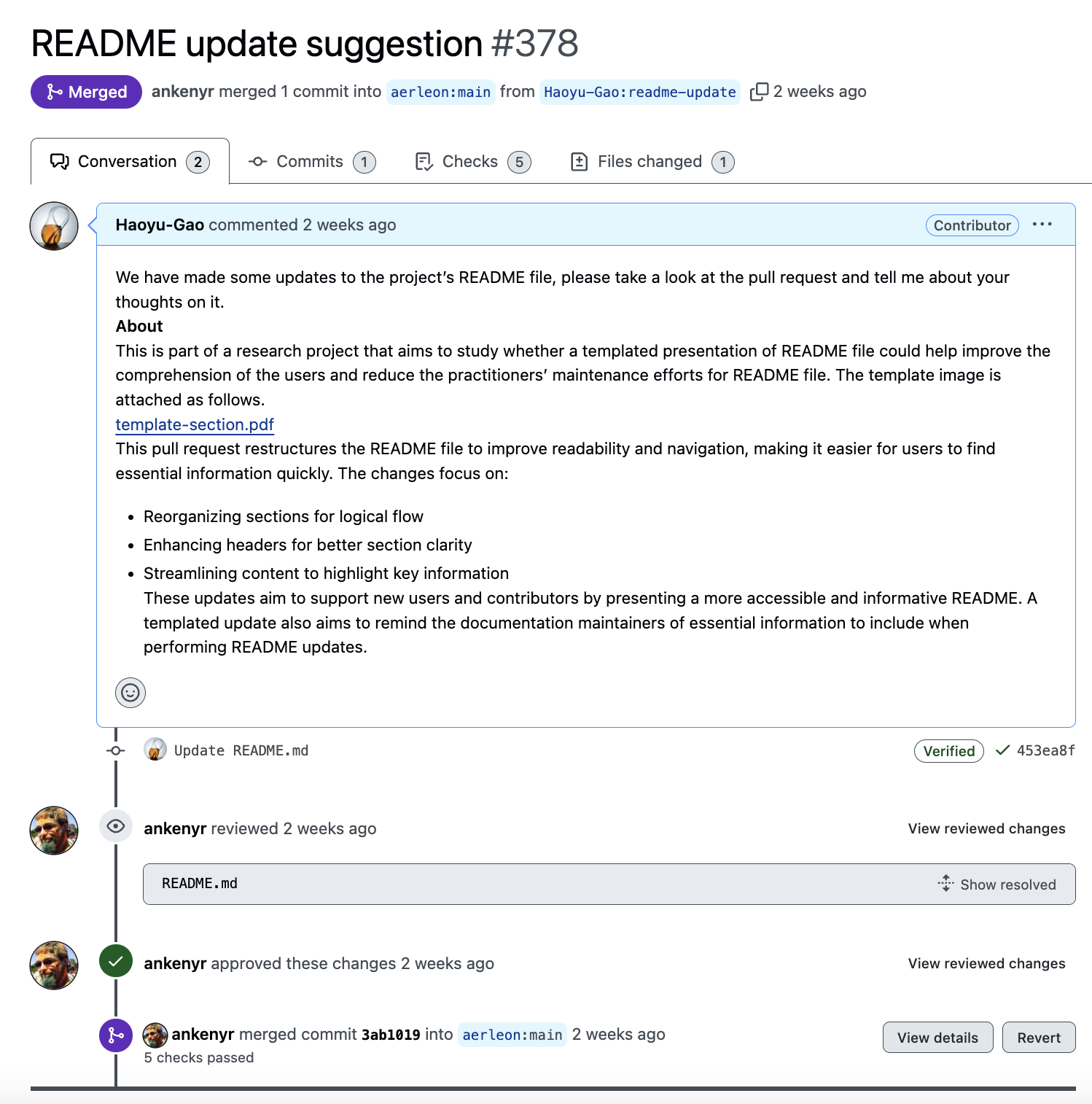}
        \caption{Example of Merged Pull Request}
        \label{fig:merged-PR}
    \end{figure}

\textbf{Positive feedback. } Figure~\ref{fig:merged-PR} shows the merged pull request to project \texttt{aerleon/aerleon}.\footnote{\url{https://github.com/aerleon/aerleon/pull/378}} Similarly, project \texttt{spring/uberserver} and \texttt{actinia-org/arctinia-core} also directly merged our proposed change without further comment.\footnote{\url{https://github.com/spring/uberserver/pull/400}} Project \texttt{apache/openmeetings} responded positively with ``I really like the new look of the readme'' and suggested some customised changes.\footnote{\url{https://github.com/apache/openmeetings/pull/196}} The pull request was then merged into the codebase after we made the suggested changes. Additionally, project \texttt{actris-cloudnet/cloudnetpy} adopted part of our proposed changes while leaving other parts of their README unchanged.\footnote{\url{https://github.com/actris-cloudnet/cloudnetpy/pull/110}} Specifically, they adjusted section header hierarchies to improve consistency. For project \texttt{vespa-engine/pyvespa},\footnote{\url{https://github.com/vespa-engine/pyvespa/pull/976}} the maintainer commented, ``\textit{I think we can add some of the suggestions, but some also are not correct/ helpful.}'', and further suggested that ``\textit{do a human review before creating a PR}''. This feedback highlights the potential for hallucinations in ChatGPT-generated suggestions and underscores the importance of human validation to ensure correctness when updating documentation.

\textbf{Negative feedback.} Out of the two negative responses, one reason for rejection is the lack of project-specific details in the updated version. For example, project \texttt{3dcitydb/importer-exporter} mentioned that ``\textit{Some sections that are important for us, like the partners and supporters and additional information about the technical standards behind the project are missing}''.\footnote{\url{https://github.com/3dcitydb/importer-exporter/pull/312}} Additionally, both repositories viewed the proposed updates as offering insignificant improvement.

Therefore, we see that our proposed method can be adopted by open-source repositories and is endorsed by several merges and positive comments from real-life projects. However, the practitioners also raised critical points regarding the completeness and correctness of the generated documentation. Semi-automated documentation updates with humans in the loop might be a solution for such concerns while reducing manual efforts to a large extent.

It is important to note that developing a directly applicable pipeline for applying the template to README files is not the primary focus of this paper. Instead, the validations obtained through user feedback and this pull request study serve as a proof of concept, demonstrating that the template has the potential to make real-world impact. Future research should focus on further refining the template and developing a comprehensive methodology for its application, ensuring it can be seamlessly integrated into real-world production workflows.

\section{Implications and Discussion}

The ultimate goal of our study is to understand the practices of updating README files in the parts related to the installation, enabling better documentation activities to improve the quality of software documentation and its process. In this section, we discuss the findings of this study and their implications for different stakeholders, including researchers and practitioners.

\textbf{Completeness considerations}. The importance of documentation content completeness issue has been identified in previous work~\cite{aghajani2019software}, and efforts have been made to use recommendation systems to recommend newly added code elements to be documented~\cite{dagenais2014using}. However, more factors constitute the completeness factor in the context of README files. We provide a more detailed discussion of completeness considerations in the context of GitHub README files.

For documentation maintainers, instruction completeness is important. The categories that emerged from our qualitative analysis of pre-installation instructions, installation-related instructions, and post-installation instructions collectively account for a complete process, enabling self-contained documentation for users to follow. In our synthesised codes, the introduction of a new instruction without a pre-existing alternative implies a deficiency in the previous documentation. For example, we found that the programming language and operating system are frequently newly added to the document. This type of update addresses previously absent details. On the other hand, practitioners should also take into account the completeness of help information. Well-managed help information can improve the comprehension of the documentation and reduce the search time for users in case of errors~\cite{chen2014asked}. Based on our analysed data, it is evident that the frequency of help information is notably lower compared to the combined occurrence of instructions within the three major categories. This observation implies a tendency among practitioners to overlook the inclusion of comprehensive help information. Therefore, practitioners could closely monitor the issues and pull requests to incorporate frequently-asked or confused questions into the README documents. For documentation updates, they could use our proposed template as a checklist before performing the updates.

For researchers, we suggest several directions for investigation to improve the completeness of the README files. First, this work primarily focuses on the update operation with less attention on the state of README file before and after modification. Therefore, identifying whether the updates are complete in reflecting the system change and fixing the previous document's lack of information would help better characterise a ``good'' or ``bad'' update. Second, an investigation of the temporal aspect of the absence of various types of instructions within the README file could better quantify the severity of this issue. Furthermore, researchers can also assist in providing essential help information.  Automatically supplying help information is a difficult task. However, this can be done by algorithms to detect insufficient information for each instruction. Moreover, previous efforts~\cite{treude2016augmenting} augmented API documentation with information from Stack Overflow. Similar ideas could be adopted by automatically extracting the most requested help information from various online channels based on certain project characteristics. Lastly, the development of tools to detect completeness issues could help document maintainers identify and address these concerns quickly.

GitHub could also help mitigate this issue by providing the template proposed in this study for installation-related instructions when users create or update the README files. According to the result discussed in the previous section, by simply applying the template to the original document, documentation readers could experience increased documentation quality. In fact, the survey participants left comments on our proposed template: ``Good to see that you provide Frequently Asked Questions (FAQ)'', ``Covers all the important elements''. Therefore, this template could serve as a good starting point for reminding practitioners of the essential aspects to cover for their README files before performing updates.

\textbf{Correctness and up-to-dateness considerations}. Issues related to correctness and up-to-dateness have been explored in prior studies~\cite{uddin2015api, wen2019large}. Tan et al.~\cite{tan2022detecting} introduced a method to identify outdated code elements in README files. In our discussion, we offer a more in-depth exploration, encompassing considerations of correctness within the context of README information and providing a more comprehensive list of up-to-dateness considerations other than code elements.

As indicated in the previous section, errors occur across different categories. Less severe errors encompass grammar issues, typographical errors within the text, and Markdown syntax errors. More severe errors are related to instructions, which require a certain level of expertise to identify. Although not explicitly coded, we identified a few updates fixing errors within their instructions. In combination, these two types of errors constitute a noticeable percentage of update behaviours. This type of documentation effort can be avoided by checking the content more carefully. 

Concerning considerations of up-to-dateness, delays in updating the README file can result in a temporary misalignment and potentially introduce correctness issues in the document. Codes related to updating various instructions reflect the changes in the software system. As also discovered in~\cite{tan2022detecting}, code elements are indeed updated frequently in our analysed data. Specifically, the frequent pattern found for code \textit{update code demo for using this project} corresponds to modifying code elements to reflect naming paradigm shifts within the source code. Additionally, change factors regarding metadata (project names, file path names, and technology used) of the project, modification of dependencies for the project, and instructions with specific operating systems, all influence the up-to-dateness of the README file and require close monitoring by document maintainers.

\textbf{Information presentation considerations}. Previous studies investigated how information is presented within the software documentation~\cite{aghajani2019software}. Considering the role of README files, which serve as the introductory point for users to participate in OSS projects, we discuss their distinctive characteristics and implications.

As indicated in Table~\ref{tab:taxonomy}, document presentation undergoes frequent updates, ranking as the second most frequently updated category in our analysis. This observation underscores the considerable effort practitioners invest in updating this particular aspect of the documentation. However, most of the updates are repetitive and could be avoided by adopting better documentation schemes. Within this category, the code \textit{change section header name} has the highest frequency of 88 occurrences, followed by codes that adjust the hierarchies of the sections and move the contents from one section to another. Therefore, practitioners can adopt more standardised section headers and examine whether there are mismatches between contents and headers before committing the change. For researchers, natural language processing techniques can be applied to examine content-header similarity or summarise the section contents with more appropriate headers~\cite{liu-lapata-2019-text}. In addition, techniques can be applied to examine and adjust the hierarchical structure for the README sections. Furthermore, this observation also motivates us to propose the README installation-related templates in Figure~\ref{fig:template-md} to mitigate practitioners' efforts in document structure modifications.

Text rephrasing is also frequently observed within README updates. Two codes dedicated to adjusting the wording within the text or making sentences succinct account for 53 occurrences in total. For document maintainers, we suggest that they consider the audience of README files before writing the document. Specifically, as README files target newcomers and serve as an introductory document, overly complicated sentences and terminologies should be avoided. For researchers, previous research~\cite{khan2021automatic, gao2023evaluating} attempted to identify documentation smells or simplify complicated sentences within the documentation. With the advent of ChatGPT~\cite{openai2023gpt4}, text rephrasing becomes easier, and attention is drawn to how different groups of users perceive useful information.

README files serve as an introductory resource for navigating the OSS projects and should not be regarded as exhaustive documentation. In our analysed data, we observed instances where the concept of project running instructions involved presenting a comprehensive list of all running options with varying parameter combinations. This approach could lead to information overload, making it challenging for documentation users to locate their desired content. In addition, we also identified a pattern of content removal in combination with external document addition, when content within the README files becomes excessively long. For example, there are a few cases where instructions are removed and a link to \texttt{INSTALL.md} or \texttt{CONTRIBUTE.md} is supplied in the README files. 

For practitioners, the above-mentioned external content organisation approach can be adopted to avoid information overload. Furthermore, we identified in code \textit{add code instruction for running this project} that add a comprehensive list of project running instructions, making it difficult for users to identify the desired content. In this scenario, these instructions can be documented in a \texttt{RUNNING.md} with README providing a reference to it, or covered by a command line interface with parameter \texttt{--help}.

For researchers, detecting information overload helps determine when a documentation refactor of moving content outside the README file is needed. A survey to documentation users on the appropriate amount of information can help better understand the users' needs, while automated tools can also be developed for overload detection.

\textbf{Automatic documentation updates.} By monitoring common open-source channels such as commit messages, pull requests, issues, and previous versions of the documentation, we identified various triggers for documentation updates. An automatic tool for suggesting documentation updates by taking this information into consideration could help developers quickly identify the need for documentation updates during the software development process. Future research could also investigate more from detailed source code level to identify additional triggers for documentation updates.

Additionally, the results in the template validation section indicate that documentation users perceive improvements in four key dimensions -- structure, understandability, conciseness, and readability -- after applying our template to the original README documents. In short, the application of this template directly mitigates the above-mentioned issues related to documentation structure, targeted audience, and information overload. In particular, one participant commented on the templates ``\textit{I like the order of chapters and it is easy to find everything. Just enough information to get started with technology.}''. In addition, our pull request study further shows that applying this template is also perceived as useful by documentation maintainers.

Therefore, to enhance the downstream users' experience, documentation maintainers could manually adopt our template, or use an automatic approach of delegating this task to LLMs. The automatic approach could further alleviate the massive efforts from practitioners to adjust documentation presentation. However, this process should incorporate a human-in-the-loop approach, involving manual review and adjustments to account for specific details, as a general template may not be suitable for all scenarios.

For researchers, our approach did not extensively adopt prompt engineering or make adjustments to the generated documentation. Refinements in these aspects might further improve the quality of README documents. The usability of the augmented documents could be verified through live installation sessions with documentation users.

Furthermore, integrating documentation update suggestion tools with our proposed template has the potential to automate documentation updates, addressing issues related to up-to-dateness, correctness, and presentation simultaneously. This pipeline could be seamlessly integrated into the GitHub pull request process, where documentation suggestions are generated based on changes in relevant channels and the status of the previous version of the README. Future research could focus on refining these tools or enhancing the template to realise this vision.

\section{Threats to Validity}
We consider threats to the validity of our study. 
  
First, we only collect repositories that use Python or Java as their primary programming language, which could affect the generalisability of our findings to other software projects. However, we did not observe any significant differences in the taxonomy between Python and Java projects in our analyzed data. The taxonomy derived in this study includes modifications in the installation process, help information, documentation presentation, and resource management. These factors can serve as a starting point for researchers and practitioners when addressing projects with different programming languages.

Second, as indicated earlier, our semi-automatic labelling heuristic is not completely accurate in distinguishing ``relevant'' and ``irrelevant'' commits. In the sampled dataset, we discarded the ones that were misclassified during our manual analysis, and filled each sample bucket with additional randomly sampled commits to ensure the same amount of data is analysed.  

Third, the 400 analysed repositories may not be representative of the overall population, potentially introducing sample bias. To address this threat, we ensured that the sampled dataset size exceeded a 95\% confidence rate with a 5\% margin of error for the overall dataset. Furthermore, our qualitative analysis may inherently carry subjectivity. To mitigate this concern, a rigorous process was implemented, involving all authors in reviewing the codes. Weekly meetings were held to facilitate discussions and address any discrepancies in codes, concepts and categories. 

Fourth, this study is conducted on the GitHub platform, and its generalisability is expected to extend to other open-source software (OSS) projects that follow similar processes. However, considering the different procedures and use cases for proprietary and private software projects, it is important to note that we do not assert generalisability beyond the domain of open source.

Fifth, our analysis is limited in README update contents, without examining external referenced links. The results may also only apply to GitHub and are not easily generalisable beyond GitHub projects to other platforms and closed-source projects.

Sixth, our procedure for identifying triggers for documentation updates may miss certain triggers, such as changes embedded in commits outside of pull requests. Additional reasons for updates might emerge from a deeper analysis of the codebase. However, the procedure was designed to monitor the primary channels within open-source repositories. This approach provides a systematic way to identify triggers without requiring exhaustive manual searches. Moreover, this design offers a foundation for automating the linkage between triggers and documentation updates, potentially enabling the identification of more triggers in the future.

Seventh, our template validation was conducted on 20 README documents with 30 participants, which might not represent the entire range of README documentation genres or the broader participant population. To mitigate this limitation, we randomly sampled the evaluated documents and recruited participants who work in IT with an additional ``quality gate''. However, the participants were not actual README users; they were only required to evaluate the perceived documentation quality without installing the software according to the guidelines provided in the documents. Additionally, no formal training of participants except the definition is provided before the survey, leading to a relatively lower agreement with some level of subjectivity. Therefore, we only claim an improvement in the documentation reading experience rather than in documentation accuracy.  Future work can further investigate this perspective by involving actual README users with live installation sessions. 

Lastly, the pull request validation study only covers 30 repositories and may not generalise to all repositories on GitHub. Moreover, this investigation does not provide long-term data about whether the merged changes will remain in the repository. Furthermore, some repositories had not responded to our pull requests at the time of paper submission, preventing us from including those results in the current version of the manuscript.



\section{Conclusion}
README files serve a critical role of introductory overviews for software projects, among which the installation-related sections are included. Considering various issues these documents may encounter brought by the rapidly evolving software repositories and ecosystem, as well as the benefits of investing efforts to maintain high-quality documentation, knowledge of the practices of the README file efforts becomes essential. To achieve this, we conducted a large-scale qualitative analysis on GitHub README file commit updates and their associated triggers specifically related to installation sections, aiming to gain comprehensive insight into documentation activities.

In this paper, we collected GitHub repositories with several stringent filters to obtain a dataset consisting of 12,908 repositories. From this dataset, we randomly sampled 400 repositories, resulting in 25,209 commits that modified a README file. We applied semi-automatic labelling heuristics and eventually obtained 8,673 commits that modified sections related to installation topics. Among these, 1,168 instances were cluster sampled for qualitative analysis, resulting in a comprehensive taxonomy of 189 update behaviours.

Our research reveals six categories of updates, namely pre-installation instructions, installation instructions, post-installation instructions, help information updates, document presentation, and external resource management. We further provide detailed insights into modification behaviours and offer examples of these updates.  We also investigated the triggers for documentation updates, which led to three categories including errors in the previous documentation, changes in the codebase, and need for documentation improvement.

Based on our findings, we proposed a README template for documentation maintainers when updating documents. The template validation process finds that the augmented documents based on our template are perceived as higher quality by documentation readers. Moreover, our pull request study also shows that documentation maintainers find it useful. This step serves as an initial step towards automated documentation updates to facilitate better documentation activities. Furthermore, we provided suggestions for practitioners and potential directions for researchers from perspectives of completeness, correctness and up-to-dateness, and information presentation considerations. To sum up in an abstract way, documentation maintainers should take care of completeness and correctness for instructions in a succinct manner, while incorporating help information and external resources to make README files more accessible and easier to comprehend. Document presentation is also a critical aspect to consider. Meanwhile, future research directions include identifying triggers for README updates, understanding user information needs in documentation, and developing automated documentation tools. Knowledge of triggering events could reveal the relationship between document and code activities, user information needs help understand the appropriate instructions to include, while all these directions contribute to the eventual automated documentation tools.

\ifCLASSOPTIONcaptionsoff
  \newpage
\fi

\bibliographystyle{IEEEtran}
\bibliography{reference}

\end{document}